\documentclass[a4paper,11pt]{article}
\pdfoutput=1 % if your are submitting a pdflatex (i.e. if you have
\usepackage{jheppub} % for details on the use of the package, please
\usepackage[T1]{fontenc} % if needed
 \usepackage{verbatim}
\def\be{\begin{equation}}
\def\ee{\end{equation}}
\def\ba{\begin{eqnarray}}\def\bea{\begin{eqnarray}}
\def\ea{\end{eqnarray}}  \def\eea{\end{eqnarray}}

\def\nn{\nonumber}

\def\a{\alpha}
\def\b{\beta}
\def\d{\delta}
\def\ep{{\epsilon}}
\def\g{\gamma}

\def\l{\lambda}
\def\O{\Omega}\def\o{\omega}
\def\th{\theta}

\def\[{\left[}
\def\]{\right]}
\def\({\left(}
\def\){\right)}

\def\<{\langle}
\def\>{\rangle}

   \def\sp2n{Sp(2N)}

\def\2F1{\,_2{\rm F}_1}

\title{\boldmath No black hole bomb for $D$-dimensional non-extremal Reissner-Nordstrom black holes against charged massive scalar perturbation}
\author[a,b]{Jia-Hui Huang} %\note{Corresponding author.}}

\affiliation[a]{Key Laboratory of Atomic and Subatomic Structure and Quantum Control (Ministry of Education), Guangdong Basic Research Center of Excellence for Structure and Fundamental Interactions of Matter, School of Physics, South China Normal University, Guangzhou 510006, China}
\affiliation[b]{Guangdong Provincial Key Laboratory of Quantum Engineering and Quantum Materials, Guangdong-Hong Kong Joint Laboratory of Quantum Matter, South China Normal University, Guangzhou 510006, China}
\emailAdd{huangjh@m.scnu.edu.cn}

\abstract{The superradiant stability of asymptotically flat $D$-dimensional non-extremal Reissner-Nordstrom black holes under charged massive scalar perturbation is analytically studied. In previous works, it proved that there are no black hole bombs for five and six-dimensional non-extremal Reissner-Nordstrom black holes against charged massive scalar perturbation. In this work, we extend the previous discussions to the $D$-dimensional case ($D\geq7$) and find that the same conclusion holds in arbitrary higher dimensional case.}

\begin{document}

\maketitle
\flushbottom
%\tableofcontents

\section{Introduction}
Black holes are important objects in both theoretical and observational physics. Linear (in)stability analysis of black holes plays an important role in many topics, such as the (in)stability of black hole solutions, the black hole ringdown phase after binary merger and astrophysics \cite{Regge:1957td,Barack:2018yly,Pani:2013pma}.
When a charged rotating black hole is scattered off  by a charged bosonic field, the electromagnetic or (and) the rotational energy of the black hole may be extracted by the external field under certain conditions. This phenomenon is called superradiance \cite{Cardoso2004,Brito:2015oca,Brito:2014wla}. In order to trigger the superradiance, the angular frequency $\omega$ of the scattered field should satisfy the following superradiance condition
  \begin{equation}\label{superRe}
   \omega < \text{m}\Omega_H  + e\Phi_H,
  \end{equation}
where $e$ and $\text{m}$ are the charge and azimuthal number of the bosonic wave mode, $\Omega_H$ is the angular velocity of the black hole horizon and $\Phi_H$ is the electromagnetic potential of the black hole horizon.
The superradiant scattering was studied long time ago \cite{P1969,Ch1970,M1972,Ya1971,Bardeen1972,Bekenstein1973,Damour:1976kh}, and has broad applications in various areas of physics(for a comprehensive review, see\cite{Brito:2015oca}). If there is a ``mirror'' between the black hole horizon and spatial infinity, 
 the amplified perturbation will be scattered back and forth between the ``mirror'' and black hole horizon, and this will lead to the superradiant instability of the system. This is the so-called black hole bomb mechanism \cite{PTbomb,Cardoso:2004nk,Herdeiro:2013pia,Degollado:2013bha}.

The superradiant (in)stability of four-dimensional rotating Kerr black holes under massive scalar or vector perturbation has been studied in \cite{Huang:2019xbu,Strafuss:2004qc,Konoplya:2006br,Cardoso:2011xi,Dolan:2012yt,Hod:2012zza,Hod:2014pza,Aliev:2014aba,Hod:2016iri,Degollado:2018ypf,Ponglertsakul:2020ufm,
East:2017ovw,East:2017mrj,Lin:2021ssw,Xu:2020fgq}.
Rotating or charged black holes with certain asymptotically curved space are proved to be superradiantly unstable under massless or massive bosonic perturbation \cite{Cardoso:2004hs,Cardoso:2013pza,Zhang:2014kna,Delice:2015zga,Aliev:2015wla,Wang:2015fgp,
Ferreira:2017tnc,Wang:2014eha,Bosch:2016vcp,Huang:2016zoz,Gonzalez:2017shu,Zhu:2014sya}, where the asymptotically curved geometries provide  natural mirror-like boundary conditions.
The four-dimensional asymptotically flat extremal or non-extremal Reissner-Nordstrom(RN) black hole has proved superradiantly stable against charged massive scalar perturbation \cite{Hod:2013eea,Huang:2015jza,Hod:2015hza,DiMenza:2014vpa,Mai:2021yny,Zou:2021mwa}. The argument is that the two conditions for the possible superradiant instability of the system, (i) existence of a trapping potential well outside the black hole horizon and (ii) superradiant amplification of the trapped modes, can't be satisfied simultaneously in the RN black hole and scalar perturbation system\cite{Hod:2013eea,Hod:2015hza}.

The linear stability of higher dimensional black holes has also been studied in the literature (for an incomplete list, see\cite{Konoplya:2011qq,Konoplya:2007jv,Konoplya:2008au,Konoplya:2013sba,Konoplya:2008rq,Kodama:2003kk,Kodama:2007sf,Ishibashi:2011ws,Ishihara:2008re,Ishibashi:2003ap,Destounis:2019hca}).
In Ref.\cite{Konoplya:2008au}, the asymptotically flat RN black holes in $D$=5,6,..,11 are shown to be stable by studying the time-domain evolution of
the massless scalar perturbation with a numerical method. In Ref.\cite{Konoplya:2013sba}, the authors provided numerical evidence that asymptotically flat extremal RN black holes are stable for arbitrary $D$ under massless perturbation. 

Recently, an analytical method based on the \textit{Descartes' rule of signs} has been used to study the superradiant stability of higher dimensional RN black holes under charged massive scalar perturbation\cite{Huang:2021dpa,Huang:2021jaz,huang2022-6d}. It proved that there is no black hole bombs for five- and six-dimensional (non-)extremal RN black holes under charged massive scalar perturbations and the system is superradiantly stable. It is also find that the above conclusion still holds for the $D$-dimensional extremal RN black hole case \cite{huang2022-d}.

In this work, we will use the above mentioned analytical method to study the superradiant stability of $D$-dimensional ($D\geq 7$) non-extremal RN black hole against charged massive scalar perturbation. We show that there is no potential well for the effective potential experienced by the scalar perturbation. The conditions for the possible black hole bomb can not be satisfied simultaneously, so there is no black hole bomb for the $D$-dimensional non-extremal RN black hole and charged massive scalar perturbation system.

The paper is organized as follows: In Section 2, we present the description of the model and the asymptotic analysis of boundary conditions. In Section 3, the effective potential of the radial equation of motion is given, and the asymptotic behaviors of the effective potential at the horizon and spatial infinity are discussed.
In Section 4, we give a brief description of the proof of our main result that there is no potential well outside the black hole horizon for the superradiant bound modes. The details of the proof are in the appendix. The final Section is devoted to the summary.

\section{Model description}
In this section, we present the model in which we are interested, i.e. a $D$-dimensional non-extremal RN black hole against charged massive scalar perturbation.
The metric of the $D$-dimensional non-extremal RN black hole \cite{Myers:1986un,Huang:2021jaz,Huang:2021dpa} is
\bea
ds^2=-f(r)dt^2+\frac{dr^2}{f(r)}+r^2d\O_{D-2}^2.
\eea
The function $f(r)$ reads
\bea
f(r)=1-\frac{2m}{r^{D-3}}+\frac{q^2}{r^{2(D-3)}},
\eea
where the parameters $m$ and $q$ are related with the ADM mass $M$ and electric charge $Q$ of the RN black hole,
\bea
m=\frac{8\pi}{(D-2)Vol(S^{D-2})}M,~~ q= \frac{8\pi}{\sqrt{2(D-2)(D-3)}Vol(S^{D-2})}Q.
\eea
Here $Vol(S^{D-2})=2\pi^{\frac{D-1}{2}}/\Gamma(\frac{D-1}{2})$ is the volume of unit ($D-2$)-sphere.
$d\O_{D-2}^2$ is the common line element of a ($D-2$)-dimensional unit sphere $ S^{D-2}$ and can be written as
\bea
d\O_{D-2}^2=d\th_{D-2}^2+\sum^{D-3}_{i=1} \prod_{j=i+1}^{D-2}\sin^2(\th_{j})d\th_i^2,
\eea
 where the ranges of the angular coordinates are taken as $\th_i\in [0,\pi](i=2,..,D-2), \th_1\in [0,2\pi]$.
The inner and outer horizons of this RN black hole  are
  \bea
  r_\pm=(m\pm\sqrt{m^2-q^2})^{1/(D-3)}.
  \eea
The event horizon of the black hole is located at $r_h= r_+$. We introduce two symbols $u, v$, defined as $u=r_+^{D-3},v=r_-^{D-3}$.  
It is obvious that we have the following two equalities:
\bea
u+v=2m,~uv=q^2.
\eea
 The electromagnetic field outside the black hole horizon is described
 by the following 1-form  vector
 \bea
 A=-\sqrt{\frac{D-2}{2(D-3)}}\frac{q}{r^{D-3}} dt=-c_D\frac{q}{r^{D-3}} dt.
 \eea

The equation of motion for a charged massive scalar perturbation in this $D$-dimensional non-extremal black hole background is governed by the covariant Klein-Gordon equantion
\bea
(D_\nu D^\nu-\mu^2)\phi=0,
\eea
where $D_\nu=\nabla_\nu-ie A_\nu$ is the covariant derivative and $\mu,~e$ are the mass and charge of the scalar field respectively.
Since the RN black hole is stationary, the solution of the above equation with definite angular frequency  can be written as
\bea
\phi(t,r,\th_i)=e^{-i\o t}R(r)\Theta(\th_i).
\eea
The angular eigenfunctions $\Theta(\th_i)$  are $(D-2)$-dimensional scalar spherical harmonics and the corresponding eigenvalues are given by $-l(l+D-3), (l=0,1,2,..)$\cite{Chodos:1983zi,Higuchi:1986wu,Rubin1984,Achour:2015zpa,Lindblom:2017maa}.

The radial equation of motion is described by
\bea\label{eq-radial}
\Delta\frac{d}{dr}(\Delta\frac{d R}{dr})+U R=0,
\eea
where
\bea\nn
\Delta&=&r^{D-2}f(r),\\
U&=&(\o+e A_t)^2 r^{2(D-2)}-l(l+D-3) r^{D-4}\Delta-\mu^2 r^{D-2}\Delta.
\eea

In order to analyze the physical boundary conditions needed here at the horizon and spatial infinity, we define the tortoise coordinate $y$ by $dy=\frac{r^{D-2}}{\Delta}dr$ and a new radial function $\tilde{R}=r^{\frac{D-2}{2}}R$, then the radial equation \eqref{eq-radial} can be rewritten as
\bea
\frac{d^2\tilde{R}}{dy^2}+\tilde{U} \tilde{R}=0,
\eea
where
\bea
\tilde{U}=\frac{U}{r^{2(D-2)}}-\frac{(D-2)f(r)[(D-4)f(r)+ 2 r f'(r)]}{4r^2}
\eea
The asymptotic behaviors of $\tilde{U}$ at the spatial infinity and the outer horizon are
\bea
\lim_{r\rightarrow +\infty}\tilde{U}= \o^2-\mu^2,~~
\lim_{r\rightarrow r_h} \tilde{U}= (\o-c_D\frac{e q}{r_h^{D-3}})^2=(\o-e\Phi_h)^2,
\eea
where $\Phi_h$ is the electric potential of the outer horizon of the RN black hole. Since we just consider a classical black hole and are interested in the black hole bomb mechanism, 
 we need purely ingoing wave condition at the horizon and bound state condition at spatial infinity, which leads to the following two conditions
\bea\label{sup-con}
\o&<&e\Phi_h=c_D\frac{e q}{r_+^{D-3}},\\
\label{bound-con}
\o&<&\mu.
\eea
The first inequality is the superradiance condition and the second inequality gives the bound state condition.

\section{Effective potential and its asymptotic behaviors}
In order to analyze the superradiant stability of the RN black hole and scalar perturbation system, we define a new radial function $\psi=\Delta^{1/2} R$, then the radial equation of motion \eqref{eq-radial} can be written as a  Schrodinger-like equation
\bea
\frac{d^2\psi}{dr^2}+(\o^2-V)\psi=0,
\eea
where $V$ is the effective potential, which is the main object we will discuss. The explicit expression for the effective potential  $V$ is $V=\o^2+\frac{B_1}{A_1}$ and
\bea\label{V}
A_1&=&4r^{2}(r^{2D-6}-2 m r^{D-3}+q^2)^2,\\
B_1&=&4(\mu^2-\o^2)r^{4D-10}+(2l+D-2)(2l+D-4)r^{4D-12}-8(m\mu^2-c_D eq\o)r^{3D-7}\nn\\
&-&4m(2\l_l+(D-4)(D-2))r^{3D-9}+4q^2(\mu^2-c_D^2 e^2)r^{2D-4}\nn\\
&-&2(2m^2-q^2(2 \l_l+3(D-4)(D-2)+2))r^{2D-6}-4m q^2(D-4)(D-2)r^{D-3}\nn\\
&+&q^4(D-4)(D-2),
\eea
where $\l_l=l(l+D-3)$.

\subsection{Asymptotic behaviors of the effective potential}

Defining $D_1=D^2-6D+8=(D-2)(D-4)$, the derivative of effective potential $V$ reads
\bea
V'(r)&=&-\frac{C(r)}{2r^3(r^{2(D-3)}-2m r^{D-3}+q^2)^3},\\
C(r)&=&a_6 r^{6(D-3)}+a'_5 r^{5D-13}+a_5 r^{5(D-3)}+a'_4 r^{4D-10}+a_4 r^{4(D-3)}\nn\\
&+&a'_3 r^{3D-7}+a_3 r^{3(D-3)}+a'_2 r^{2D-4}+a_2 r^{2(D-3)}+a_1 r^{D-3}+a_0,
\eea
where
\bea\label{D-C-xishu}
&&a_6=4\l_l+D_1, a'_5=4(D-3)(c_D eq\o+m\mu^2-2m\o^2),\nn\\
&&a_5=2m[(2D-14)\l_l-3D_1],\nn\\
&& a'_4=-4(D-3)[2m^2\mu^2-2c_D m eq\o+q^2(\mu^2-2\o^2+c_D^2 e^2)],\nn\\
&&a_4=-[4m^2((2D-10)\l_l+(D-2)(D^2-9D+21))\nn\\
&&~~~~~~~~~+q^2((4D-20)\l_l-(D-2)(4D^2-21D+24))]\nn\\
&&a'_3=12(D-3)(m\mu^2-c_D eq\o)q^2, a'_2=4(D-3)(c_D^2 e^2-\mu^2)q^4,\nn\\
&&a_3=[8m^3+4mq^2((3D-13)\l_l-5(D-2)(D-4)-2)]\nn\\
&&a_2=(D-4)q^2[(4D^2-12D+12)m^2-(4\l_l+4D^2-27D+42)q^2],\nn\\
&& a_1=-6(D-2)(D-4)mq^4, a_0=(D-2)(D-4)q^6
\eea

When $D\geq 7 $, the effective potential $V$ and its derivative $V'(r)$ have the following  asymptotic behaviors,
\bea\nn
&&r\rightarrow r_h,~~ V\rightarrow -\infty,\\\nn
&&r\rightarrow +\infty,~~ V\rightarrow \mu^2,\\
&&r\rightarrow +\infty,~~V'(r)\rightarrow -\frac{(2l+D-2)(2l+D-4)}{2r^3}<0.
\eea
From the above asymptotic behaviours, we conclude that the effective potential $V(r)$ has at least one maximum outside the event horizon $r_h$ and there is no potential well near the spatial infinity. In the next section, we will prove that in fact there is no potential well between the event horizon and spatial infinity for superradiant bound states by analyzing the derivative of the effective potential $V$.

\section{Further analysis of the effective potential $V$}
In this section, we show that there is only one extreme for the effective potential outside the
RN black hole horizon by analyzing the real roots of the derivative of the effective potential.
Explicitly, it is shown that only one real root exists for the following equation
\bea
V'(r)=0
\eea
when $r > r_h$. Since we are interested in the real roots of $V'(r)$, we just consider the numerator of $V'$, i.e. $C(r)$, which is a polynomial of $r$. 

Making a change of variable from $r$ to $z=r-r_h$, $C(r)$ is changed to a polynomial of $z$, $C(z+r_h)$, which can be expanded as
\bea\label{bp}
C(z+r_h)=\sum_{i=0}^{6D-18} b_i z^i.
\eea
After the change of variables, a real root for $C(r)=0$ with $r > r_h$ corresponds to a positive real root for $C(z+r_h)=0$ with $z>0$. 
In order to analyze the number of positive real roots for $C(z+r_h)=0$, we will use a method based on 
 \textit{Descartes' rule of signs}, i.e. we will consider the sign changes of the sequence of the following coefficients 
\bea
b_{6D-18},~b_{6D-19},~b_{6D-20},...,b_2,~b_1,~b_0.
\eea

Remember that $u=r_+^{D-3},v=r_-^{D-3}$ and they satisfy $u+v=2m, u v=q^2$. $u,v$ will be often used in the later discussion. 
The constant term in $C(z+r_h)$ is
\bea
b_0&=&a_0+a_1 r_h^{D-3}+a_2 r_h^{2D-6}+a'_2 r_h^{2D-4}+a_3 r_h^{3D-9}\nn\\&&+a'_3 r_h^{3D-7}+a_4 r_h^{4D-12}+a'_4 r_h^{4D-10}+a_5 r_h^{5D-15}+a'_5 r_h^{5D-13}+a_6 r_h^{6D-18}.
\eea
Plugging Eqs.\eqref{D-C-xishu} into the above equation, we can obtain
\bea
b_0=-(D-3)u^3 (u-v)[(D-3)^2(u-v)^2+4r_h^2(c_D e q-u\o)^2]<0.
\eea

It is easy to see that $a_6=D_1 + 4 \l_l>0$.  Considering the two conditions Eq.\eqref{sup-con} and Eq.\eqref{bound-con}, it is also easy to prove
\bea
a'_5&=&4(D-3)(c_D e q \o+m\mu^2-2m\o^2)>0.
\eea
So we can immediately obtain that
\bea\label{6d-5d-ptv}
b_{6D-18}&=&a_6>0,\nn\\
b_{6D-19}&=&a_{6}C_{6D-18}^{6D-19} r_h>0,\nn\\
...,\nn\\
b_{5D-12}&=&a_6 C_{6D-18}^{5D-12} r_h^{D-6}>0,\nn\\
b_{5D-13}&=&a_6 C_{6D-18}^{5D-13}r_h^{D-5}+a'_5>0,\nn\\
b_{5D-14}&=&a_6 C_{6D-18}^{5D-14}r_h^{D-4}+a'_5C_{5D-13}^{5D-14}r_h>0.
\eea

Let's see the coefficient $b_{k}$ $(4D-9\leq k\leq 5D-15)$, which is
\bea
b_{k}=a_6 C_{6D-18}^{k}r_h^{6D-18-k}+a'_5 C_{5D-13}^{k}r_h^{5D-13-k}+a_5 C_{5D-15}^{k}r_h^{5D-15-k}.
\eea
The term involving $a'_5$ is positive. Next, we prove the sum of the left two terms are also positive. 
This sum can be written as
\bea
&&a_6 C_{6D-18}^{k}r_h^{6D-18-k}+a_5 C_{5D-15}^{k}r_h^{5D-15-k}\\
&=&C_{5D-15}^{k}r_h^{5D-15-k}(a_6 u \frac{C_{6D-18}^{k}}{C_{5D-15}^{k}}+a_5 )\nn\\
&=&C_{5D-15}^{k}r_h^{5D-15-k}[D_1(-3(u+v)+u\frac{C_{6D-18}^{k}}{C_{5D-15}^{k}})+\l_l(2(D-7)(u+v)+4u\frac{C_{6D-18}^{k}}{C_{5D-15}^{k}})]\nn 
\eea
The $\l_l$ term in the sum is obviously positive when $D\geq7$. Since
\bea
\frac{C_{6D-18}^{k}}{C_{5D-15}^{k}}=\frac{6D-18}{5D-15}\frac{6D-19}{5D-16}\cdots\frac{6D-18-k+1}{5D-15-k+1}>\left(\frac{6}{5}\right)^k>
\left(\frac{6}{5}\right)^{19}>31
\eea
and $31u>6u>3(u+v)$, the $D_1$ term in the above sum is also positive.
So we have
\bea\label{4d-5d-ptv}
b_{k}>0,~(4D-9\leq k\leq 5D-15).
\eea
The signs of the other coefficients are not so easy to prove as the above. Then, we will compare the signs of two adjacent coefficients. For example,
since $b_{4D-9}>0$, when $b_{4D-10}>0$, we have
\bea
\text{sign}(b_{4D-9})=\text{sign}(b_{4D-10}),\nn
\eea
and when $b_{4D-10}<0$, we have
\bea
\text{sign}(b_{4D-9})>\text{sign}(b_{4D-10}).\nn
\eea
Thus, we have
 \bea\label{sign-4d-10}
\text{sign}(b_{p+1})\geq \text{sign}(b_{p}),~(p=4D-10).
\eea

In fact, it is hard to find certain feature of sign relations between adjacent coefficients by directly computing their difference. 
However, it is found that  after properly normalizing the coefficients with positive factors, we can obtain a feature of the sign relations of the coefficients.  
With the results in Eqs.\eqref{sign-4d-10}\eqref{sign-d-4}\eqref{sign-d-2-2d-5}\eqref{sign-d-2-d-3}\eqref{sign-2d-5-2d-6}\eqref{sign-2d-4-2d-5}\eqref{sign-2d-3-3d-10}\eqref{sign-2d-3-2d-4}
\eqref{sign-3d-8-3d-9}\eqref{sign-3d-7-3d-8}\eqref{sign-3d-6-3d-7}\eqref{sign-3d-6-4d-13}\eqref{sign-4d-12}\eqref{sign-4d-11}, we prove that
\bea
\text{sign}(b_{p+1})\geq \text{sign}(b_{p}),~(1\leq p\leq4D-10).
\eea
Together with the results that $b_0<0$ and $b_{k}>0~(4D-9\leq k\leq 6D-18)$, we find that the sign change for the sequence
$
(b_{6D-18},~b_{6D-19},~b_{6D-20},...,b_2,~b_1,~b_0)
$
is always 1. According to Descartes' rule of signs, there is at most one positive real root to the equation $C(z+r_h)=0$, i.e. there is at most one extreme for the 
effective potential outside the horizon $r_h$. And, we already know that there is at least one maximum for the effective potential from the asymptotic analysis of the effective potential. Thus, there is only a potential barrier  outside the event horizon and no potential well exists for the superradiant bound modes.

\section{Summary}

In this work, superradiant stability of  $D$-dimensional ($D\geq 7$) non-extremal RN black hole  under charged massive scalar perturbation is studied analytically. Based on
the asymptotic analysis of the effective potential $V(r)$ experienced by the scalar perturbation, it is known that there is at least one maximum for the effective potential outside the black hole event horizon. Then, we analyze the numerator of the derivative of the effective potential $C(z+r_h)$, which is a polynomial of $z=r-r_h$. We find that
\bea
b_0<0,~~b_{k}>0~(4D-9\leq k\leq 6D-18).
\eea
We also proves that
\bea
\text{sign}(b_{p+1})\geq \text{sign}(b_{p})~~(1\leq p\leq4D-10).
\eea
So the sign change in the following sequence of the real coefficients of $C(z+r_h)$
\bea
b_{6D-18},b_{6D-19},...,b_{p+1},b_p,..,b_1,b_0,
\eea
is always 1. Then according to Descartes' rule of signs, we know there is at most 1 positive root for the equation $C(z+r_h)=0$(i.e. $V'(r)=0$ when $r>r_h$). Thus, there is only one maximum for the effective potential and there is no potential well outside the horizon for the superradiant bound modes. Namely, there is no black hole bomb for $D$-dimensional non-extremal RN black hole against charged massive scalar perturbation.

\acknowledgments
This work is partially supported by Guangdong Major Project of Basic and Applied Basic Research (No. 2020B0301030008).

\appendix

\section{$\text{sign}(b_{p+1})\geq \text{sign}(b_{p})$, $ 0< p \leq D-4 $}
\label{sec-p-d-3}

An frequently used identity in the proof is
\bea\label{comb-1}
C_n^{m+1}=\frac{n-m}{m+1}C_n^m.
\eea
When $ 0 < p\leq D-3$, the coefficient $b_p$ of $z^p$ in Eq.\eqref{bp} is 
\bea\label{b-2d6d2}
b_p&=&a_6 C_{6D-18}^{p}r_h^{6D-18-p}+ a_5 C_{5D-15}^{p}r_h^{5D-15-p}\nn\\
&&+a_4 C_{4D-12}^{p}r_h^{4D-12-p}+a_3 C_{3D-9}^{p}r_h^{3D-9-p}+a_2 C_{2D-6}^{p}r_h^{2D-6-p}+a_1 C_{D-3}^{p}r_h^{D-3-p}\nn\\
&&+a'_5 C_{5D-13}^{p}r_h^{5D-13-p}+a'_4 C_{4D-10}^{p}r_h^{4D-10-p}+a'_3C_{3D-7}^{p}r_h^{3D-7-p}+a'_2C_{2D-4}^{p}r_h^{2D-4-p}\nn\\
&=&r_h^{-p}(a_6 u^6 C_{6D-18}^{p}+ a_5 u^5 C_{5D-15}^{p}+a_4u^4 C_{4D-12}^{p}+a_3 u^3C_{3D-9}^{p}+a_2 u^2 C_{2D-6}^{p}+a_1 u C_{D-3}^{p})\nn\\
&&+r_h^{-p+2}u^2(a'_5u^3 C_{5D-13}^{p}+a'_4 u^2 C_{4D-10}^{p}+a'_3 u C_{3D-7}^{p}+a'_2C_{2D-4}^{p}).\\
&=&r_h^{-p}\sum_{i=1}^{6}a_i u^i C_{i(D-3)}^p+r_h^{-p+2}u^2\sum_{j=2}^{5}a'_j u^{j-2} C_{j(D-3)+2}^p\\
&=&r_h^{-p}\check{A}_p+r_h^{-p+2}u^2 \bar{B}_p.
\eea
The normalized coefficient is defined as
\bea
b'_p=\frac{r_h^p}{\bar{f}_p}b_p=\frac{\check{A}_p}{\bar{f}_p}+r_h^2u^2\frac{\bar{B}_p}{\bar{f}_p}
=\frac{\check{A}^{L}_p}{\bar{f}_p}\l_l+\frac{\check{A}^R_p}{\bar{f}_p}+r_h^2u^2(\frac{\bar{B}^M_p}{\bar{f}_p}(\mu^2-\o^2)+\frac{\bar{B}^R_p}{\bar{f}_p}),
\eea
where $\check{A}^{L}_p$ is the coefficient of $\l_l$ in $\check{A}_p$, $\check{A}^R_p$ denotes the remaining terms in $\check{A}_p$, 
$\bar{B}^M_p$ is the coefficient of $(\mu^2-\o^2)$ in $\bar{B}_p$, $\bar{B}^R_p$ denotes the remaining terms in $\bar{B}_p$, and
\bea\nn
&&\bar{f}_p=u(u+v)C_{5D-13}^{p}+(u^2+v^2)C_{4D-10}^{p}-3v(u+v)C_{3D-7}^{p}+2v^2C_{2D-4}^{p}.\\\nn
&&\check{A}^{L}_p=2u^4\left(2u^2 C_{6D-18}^{p}+(D-7)u(u+v) C_{5D-15}^{p}-(D-5)(u^2+4uv+v^2) C_{4D-12}^{p}\right.\\\nn
&&~~~~ \left.+(3D-13)v(u+v) C_{3D-9}^{p}-(2D-8)v^2 C_{2D-6}^{p} \right)\\\nn
&&\check{A}^{R}_p=u^3\left((D-2)(D-4)u^3C_{6D-18}^p-3(D-2)(D-4)u^2(u+v)C_{5D-15}^{p} \right.\\\nn
&&~~~~\left.+u(( 4 D^3- 29 D^2 + 66 D-48)uv-(D^3-11 D^2+ 39 D-42)(u+v)^2)C_{4D-12}^{p}\right.\\\nn
&&~~~~\left.+(u + v)((u + v)^2-2(5D^2-30D+42) u v)C_{3D-9}^{p}\right.\\\nn
&&~~~~\left.+(D-4)v((D^2-3D+3)(u + v)^2-(4D^2-27D+42)uv)C_{2D-6}^{p}\right.\\\nn
&&~~~~\left.-3(D-2)(D-4)(u+v)v^2C_{D-3}^{p}\right),\nn\\
&&\bar{B}^M_p=2 u^2(D-3)\left(C_{5D-13}^p u (u + v) - C_{4D-10}^p (u^2 + 4 u v + v^2)\right.\nn\\
&&~~~~~\left.+3 C_{3D-7}^p (u + v)v-2 C_{2D-4}^p v^2\right),\nn\\
&&\bar{B}^R_p=2 (D-3) u \left(-u\bar{f}_p \o^2+2c_D e q u(C_{5D-13}^p u+C_{4D-10}^p(u+v)-3C_{3D-7}^p v)\o\right.\nn\\
&&~~~~~\left.-2c_D^2 e^2 q^2(C_{4D-10}^p u-C_{2D-4}^p v)\right).
\eea

Now we prove the normalization factor is positive, i.e.
\bea\label{bf-positive}
\bar{f}_p>0~(p\geq1).
\eea 
When $p=1$,
\bea\nn
\bar{f}_1&=&(9D-23) u^2 - 4 (D-2) u v - (D-3) v^2\\\nn
&>&(9D-23- 4 (D-2))uv- (D-3) v^2>(4D-12)v^2>0.
\eea
 When $p\geq2$,
\bea\nn
\bar{f}_p&=&u(u+v)C_{5D-13}^{p}+(u^2+v^2)C_{4D-10}^{p}-3v(u+v)C_{3D-7}^{p}+2v^2C_{2D-4}^{p}\\\nn
&>&(\frac{C_{5D-13}^{p}+C_{4D-10}^{p}}{C_{3D-7}^{p}}-3)v(u+v)C_{3D-7}^{p}>1.2v(u+v)C_{3D-7}^{p}>0.
\eea
Thus, we finish the proof of Eq.\eqref{bf-positive}.

In the following four subsections, we will prove $ b'_{p+1}> b'_{p},~(0<p\leq D-4)$, and obtain
\bea\label{sign-d-4}
\text{sign}(b_{p+1})\geq \text{sign}(b_{p}), (0<p\leq D-4).
\eea
This will be achieved by respectively proving
\bea
\frac{\check{A}^{L}_{p+1}}{\bar{f}_{p+1}}>\frac{\check{A}^{L}_p}{\bar{f}_p},~\frac{\check{A}^{R}_{p+1}}{\bar{f}_{p+1}}>\frac{\check{A}^{R}_p}{\bar{f}_p},~
\frac{\bar{B}^{M}_{p+1}}{\bar{f}_{p+1}}>\frac{\bar{B}^{M}_p}{\bar{f}_p},~
\frac{\bar{B}^{R}_{p+1}}{\bar{f}_{p+1}}>\frac{\bar{B}^{R}_p}{\bar{f}_p}.
\eea

\subsection{Proof of $\frac{\check{A}^L_{p+1}}{\bar{f}_{p+1}}>\frac{\check{A}^L_p}{\bar{f}_p}$}
\label{sec-cal}

We first prove 
\bea\label{cAlam-positive}
\check{A}^{L}_p>0~(p\geq1).
\eea 
When $p=1,2,3,4$, the explicit expressions of $\check{A}^{L}_p$ are
\bea\nn
\check{A}^{L}_1&=&2 (D-3)^2 u^4 (u - v)^2,~\check{A}^{L}_2=3 (D-3)^2 u^4 (u - v) ((3D-8) u - (D-2) v),\\\nn
\check{A}^{L}_3&=&\frac{(D-3)^2 u^4}{3} ((440 + D (-333 + 61 D)) u^2 -10 (25 + D (-24 + 5 D)) u v + (-82 + D (21 + D)) v^2),\\\nn
\check{A}^{L}_4&=&\frac{(D-3)^2 u^4}{12} ((-7000 + D (8274 + D (-3083 + 369 D))) u^2\\&& -4 (370 + D (228 + D (-232 + 39 D))) u v
+ (3728 + D (-2682 + (619 - 45 D) D)) v^2)
\eea
Since $u>v>0$, it is easy to prove directly they are positive when $D\geq7$.  When $p\geq5$, it is easy to see that the sum of $C_{3D-9}^{p}$ term and $C_{2D-6}^{p}$ term in $\check{A}^{L}_p$ are positive, i.e.
\bea
(3D-13)v(u+v) C_{3D-9}^{p}-(2D-8)v^2 C_{2D-6}^{p}>0.
\eea
Since $2u^2>u(u+v)$ and $u^2+4uv+v^2<3u(u+v)$, for other terms in $\check{A}^{L}_p$, we have
\bea
&&2u^2 C_{6D-18}^{p}+(D-7)u(u+v) C_{5D-15}^{p}-(D-5)(u^2+4uv+v^2) C_{4D-12}^{p}\nn\\
&>&\left(\frac{C_{6D-18}^{p}+(D-7)C_{5D-15}^{p}}{C_{4D-12}^{p}}-3(D-5)\right)u(u+v)C_{4D-12}^{p}\nn\\
&>&(6/4)^5+(5/4)^5 (D - 7) - 3 (D - 5)>0.05D+1.23>0.
\eea
Thus, $\check{A}^{L}_p>0$ when $p\geq1$.

Now we prove
\bea\label{cALam-final}
\frac{\check{A}^L_{p+1}}{\bar{f}_{p+1}}>\frac{\check{A}^L_p}{\bar{f}_p}~~(p\geq1).
\eea
For $p=1$, since $u>v>0$ and $D\geq7$ it is easy to see 
\bea\label{cALp1}
\frac{\check{A}^L_2}{\bar{f}_2}-\frac{\check{A}^L_1}{\bar{f}_1}&=&\frac{4(D-3)^2}{\bar{f}_2\bar{f}_1} u (u - 
   v) \times\nn\\
&&[(5 D-13)(( 2 D-5) u^2- (D-1) u v) + (D-2) (D-1) v^2]\nn\\
&&>0.
\eea
For $p=2$,
\bea
\frac{\check{A}^L_3}{\bar{f}_3}-\frac{\check{A}^L_2}{\bar{f}_2}=\frac{(D-3)^2}{2\bar{f}_3\bar{f}_2}(\a_0 u^4+\a_1 u^3v+\a_2 u^2v^2+\a_3u v^3+\a_4v^4),
\eea
where
\bea
&&\a_0=400 D^4-4137 D^3+15730 D^2-25977 D+15640,\nn\\
&&\a_1=-150 D^4+762 D^3+954 D^2-9888 D+12714,\nn\\
&&\a_2=-36 D^4+888 D^3-5736 D^2+14214 D-12138,\nn\\
&&\a_3=2 D^4-6 D^3-166 D^2+924 D-1258,\nn\\
&&\a_4=-27 D^3+234 D^2-657 D+594.
\eea
Since $u>v>0$, 
\bea
&&\a_0u^4+\a_1u^3v+\a_2 u^2v^2+\a_3u v^3+\a_4v^4\nn\\
&&>(\a_0+\a_1)u^3v+\a_2 u^2v^2+\a_3u v^3+\a_4v^4\nn\\
&&>(\a_0+\a_1+\a_2)u^2v^2+\a_3u v^3+\a_4v^4\nn\\
&&>(\a_0+\a_1+\a_2+\a_3)u v^3+\a_4v^4\nn\\
&&>(\a_0+\a_1+\a_2+\a_3+\a_4)v^4>0.
\eea
In the above inequalities, we use the facts that $\a_0>0,\a_0+\a_1>0,\a_0+\a_1+\a_2>0,\a_0+\a_1+\a_2+\a_3>0,\a_0+\a_1+\a_2+\a_3+\a_4>0$ for $D\geq7$, which can be checked directly.
So we have
\bea\label{cALp2}
\frac{\check{A}^L_3}{\bar{f}_3}-\frac{\check{A}^L_2}{\bar{f}_2}>0.
\eea
In the same way, we can also explicitly prove that
\bea\label{cALp6}
\frac{\check{A}^L_{p+1}}{\bar{f}_{p+1}}-\frac{\check{A}^L_p}{\bar{f}_p}>0~~(p=3,4,5,6).
\eea
When $p\geq7$, we first rewrite $\frac{\check{A}^L_p}{\bar{f}_p}$ as $\frac{\check{A}^L_p}{\bar{f}_p}=\frac{\check{A}^L_p/C_{5D-13}^{p}}{\bar{f}_p/C_{5D-13}^{p}}$. Then, using the Eq.\eqref{comb-1}, we have
\bea\label{bfmc55}
&&\frac{\bar{f}_{p}}{C_{5D-13}^{p}}-\frac{\bar{f}_{p+1}}{C_{5D-13}^{p+1}}\nn\\
&=&\frac{D-3}{(5D-13-p)C_{5D-13}^{p}}(C_{4D-10}^p (u^2 + v^2)- 6 C_{3D-7}^p  (u + v)v +6 C_{2D-4}^p v^2)\nn\\
&>&\frac{D-3}{(5D-13-p)C_{5D-13}^{p}}((C_{4D-10}^p- 6 C_{3D-7}^p) (u + v)v +6 C_{2D-4}^p v^2)>0,
\eea
where we use the fact $\frac{C_{4D-10}^p}{ C_{3D-7}^p}>7.4$ for $p\geq7$ in the last inequality. And, we also have 
\bea\label{cALmc55}
\frac{\check{A}^L_{p+1}}{C_{5D-13}^{p+1}}-\frac{\check{A}^L_p}{C_{5D-13}^{p}}&=&\frac{2u^4}{(5D-13-p)C_{5D-13}^{p}}(\b_0 u^2+\b_1u v+\b_2 v^2)\nn\\
&>&\frac{2u^4}{(5D-13-p)C_{5D-13}^{p}}((\b_0+\b_1)u v+\b_2 v^2)>0,
\eea
where
\bea\label{beta012}
\b_0&=&C_{4D-12}^p (D-5) (D-1)+2 C_{6D-18}^p (D-5)-2 C_{5D-15}^p (D-7),\\
\b_1&=&-2 C_{3D-9}^p (D-2) (3 D-13)+4 C_{4D-12}^p (D-5) (D-1)-2 C_{5D-15}^p (D-7),\nn\\
\b_2&=&2 C_{2D-6}^p (D-4) (3 D-7)-2 C_{3D-9}^p (D-2) (3 D-13)+C_{4D-12}^p (D-5) (D-1).\nn
\eea
In the above inequalities, we use the facts that $\b_0>0$, $\b_2>0$, $\b_0+\b_1>0$ since $\frac{C_{4D-12}^p}{ C_{3D-9}^p}>7.4$ and $\frac{C_{6D-18}^p}{ C_{5D-15}^p}>3.5$ for $p\geq7$.
With the results in Eqs.\eqref{bfmc55}\eqref{cALmc55}, we conclude that
\bea\label{cALp7}
\frac{\check{A}^L_{p+1}}{\bar{f}_{p+1}}=\frac{\check{A}^L_{p+1}/C_{5D-13}^{p+1}}{\bar{f}_{p+1}/C_{5D-13}^{p+1}}>\frac{\check{A}^L_p/C_{5D-13}^{p}}{\bar{f}_p/C_{5D-13}^{p}}
=\frac{\check{A}^L_p}{\bar{f}_p}~(p\geq7).
\eea
Thus,  we finish the proof of Eq.\eqref{cALam-final} with the results in Eqs.\eqref{cALp1}\eqref{cALp2} \eqref{cALp6}\eqref{cALp7}.

\subsection{Proof of $\frac{\check{A}^R_{p+1}}{\bar{f}_{p+1}}>\frac{\check{A}^R_p}{\bar{f}_p}$}
\label{sec-car}
In this subsection, we prove 
\bea\label{cAR-final}
\frac{\check{A}^R_{p+1}}{\bar{f}_{p+1}}-\frac{\check{A}^R_p}{\bar{f}_p}>0~~(p\geq1).
\eea
We rewrite $\bar{f}_p$ as
\bea
\bar{f}_p=y_5+y_4-y_3+y_2,
\eea
where
\bea
&&y_5=u(u+v)C_{5D-13}^{p},~y_4=(u^2+v^2)C_{4D-10}^{p},\nn\\
&&y_3=3v(u+v)C_{3D-7}^{p},~y_2=2v^2C_{2D-4}^{p},
\eea
and rewrite $\check{A}^R_{p}$ as
\bea\label{car-gi}
\check{A}^R_{p}=u^3\sum_{i=1}^{6}g_i C_{i(D-3)}^p,
\eea
where
\bea
&&g_6=(D-2)(D-4)u^3,g_5=-3(D-2)(D-4)u^2(u+v),\nn\\
&&g_4=-(D-2)^3u(u-v)^2+F_2(u+v)^2u-F_1u^2v,\nn\\
&&g_3=(u-v)^2(u+v)-10(D-2)(D-4)(u+v)uv,\nn\\
&&g_2=(D-2)^2(D-4)v(u-v)^2+(D-4)v(G_1 u v + G_2 (u + v)^2),\nn\\
&&g_1=-3(D-2)(D-4)(u+v)v^2,
\eea
and
\bea
&&F_1=5D^2-18D+16,~F_2=5D^2-27D+34,\nn\\
&&G_1=11D-26,~G_2=D-1.
\eea
Now consider $g_1$ term and let's prove 
\bea\label{g1}
\frac{u^3 g_1 C_{D-3}^{p+1}}{\bar{f}_{p+1}}>\frac{u^3 g_1 C_{D-3}^{p}}{\bar{f}_p},
\eea
i.e.
\bea
\frac{C_{D-3}^{p+1}}{\bar{f}_{p+1}}<\frac{C_{D-3}^{p}}{\bar{f}_p} \Leftrightarrow (D-3-p)\bar{f}_p<(p+1)\bar{f}_{p+1}\nn\\
\Leftrightarrow (4D-10)y_5+(3D-7)y_4-2(D-2)y_3+(D-1)y_2>0.
\eea
Since $y_5/y_3>1.5/3,~y_4/y_3>1.2/3$, the above inequality holds.

Similarly, let's consider the $(D-2)(D-4)$ term in $g_3$, and prove 
\bea\nn\label{g3-1}
u^3\frac{-10(D-2)(D-4)(u+v)uvC_{3D-9}^{p+1}}{\bar{f}_{p+1}}>u^3\frac{-10(D-2)(D-4)(u+v)uvC_{3D-9}^{p}}{\bar{f}_p},\\
\eea
i.e.
\bea
\frac{C_{3D-9}^{p+1}}{\bar{f}_{p+1}}<\frac{C_{3D-9}^{p}}{\bar{f}_p} \Leftrightarrow (3D-9-p)\bar{f}_p<(p+1)\bar{f}_{p+1}\nn\\
\Leftrightarrow (2D-4)y_5+(D-1)y_4-2y_3-(D-5)y_2>0.
\eea
Since $y_5/y_3>1.5/3,~y_4/y_2>1$, the above inequality holds. 
Then, let's consider the $F_1$ term in $g_4$ and prove 
\bea\label{F1}
u^3\frac{-F_1 u^2vC_{4D-12}^{p+1}}{\bar{f}_{p+1}}>u^3\frac{-F_1u^2vC_{4D-12}^{p}}{\bar{f}_p},
\eea
i.e.
\bea
\frac{C_{4D-12}^{p+1}}{\bar{f}_{p+1}}<\frac{C_{4D-12}^{p}}{\bar{f}_p} \Leftrightarrow (4D-12-p)\bar{f}_p<(p+1)\bar{f}_{p+1}\nn\\
\Leftrightarrow (D-1)y_5+2y_4+(D-5)y_3-2(D-4)y_2>0.
\eea
Since $y_3/y_2>3,~y_5>y_2$, the above inequality holds.

Next, consider the sum of the $(u-v)^2$ terms in $g_4,~g_2$ and for $p\geq3$ let's prove
\bea\label{d3}
&&u^3\frac{ -(D-2)^3u(u-v)^2 C_{4D-12}^{p+1}+(D-2)^2(D-4)v(u-v)^2C_{2D-6}^{p+1}}{\bar{f}_{p+1}}\nn\\
&>&u^3\frac{ -(D-2)^3u(u-v)^2 C_{4D-12}^{p}+(D-2)^2(D-4)v(u-v)^2C_{2D-6}^{p}}{\bar{f}_p},
\eea
i.e.
\bea
&&\frac{ -(D-2)u C_{4D-12}^{p+1}+(D-4)v C_{2D-6}^{p+1}}{\bar{f}_{p+1}}
>\frac{ -(D-2)u C_{4D-12}^{p}+(D-4)v C_{2D-6}^{p}}{\bar{f}_p}\nn\\
&&\Leftrightarrow
C_{4D-12}^{p} [-2 (D-4) y_2 + (D-5) y_3 + 2 y_4+(D-1) y_5]\nn\\
&&>C_{2D-6}^{p} [2 y_2 -( D-1) y_3+ (2 D-4) y_4 + (3D-7) y_5].
\eea
Since $y_3/y_2>3$, the sum of $y_4,y_3,y_2$ terms on the left-hand side is positive and the sum of $y_3,y_2$ terms on the right-hand side is negative. 
When $p\geq3$, $C_{4D-12}^{p}/C_{2D-6}^{p}>8$ and the $y_5$ term on the left-hand side is greater than the sum of $y_4,y_5$ terms on the right-hand side. 
Thus, we finish the proof of Eq.\eqref{d3}.

Finally, let's consider all other terms in $\check{A}^R_{p}$, which is 
\bea
&&S_p=u^3\left(g_6C_{6D-18}^p+g_5C_{5D-15}^p+F_2(u+v)^2uC_{4D-12}^p\right.\nn\\
&&\left.+(u-v)^2(u+v)C_{3D-9}^p+(D-4)v(G_1 u v + G_2 (u + v)^2)C_{2D-6}^p\right).
\eea
The second line of $S_p$ is obviously positive. Then we prove the first line of $S_p$ is positive for $p\geq7$ in the following.
\bea
&&g_6C_{6D-18}^p+g_5C_{5D-15}^p+F_2(u+v)^2uC_{4D-12}^p>0\nn\\
&\Leftrightarrow& (D-2)(D-4)C_{6D-18}^pu^2-3(D-2)(D-4)C_{5D-15}^pu(u+v)+F_2C_{4D-12}^p(u+v)^2\nn\\
&&>0.
\eea
It is known that $u+v=2m$. To simplify the expression in following proof, hereafter, we take the mass of the black hole to be $1/2$, i.e.,$u+v=1$. 
This simplification just leads to a difference of an overall positive factor, and doesn't change the positivity or negativity of a homogeneous expression of $u,v$.  
 Replacing $u$ with $1-v$, the above inequality is equivalent to
\bea
\eta_2 v^2+\eta_1 v+\eta_0>0,
\eea
where
\bea
&&\eta_2=(D-2)(D-4)C_{6D-18}^p,\eta_1=(D-2)(D-4)(3C_{5D-15}^p-2C_{6D-18}^p),\nn\\
&&\eta_0=(D-2)(D-4)(C_{6D-18}^p-3C_{5D-15}^p)+F_2 C_{4D-12}^p.
\eea
When $p\geq7$, $C_{6D-18}^p/C_{5D-15}^p>3$, so $\eta_0>0$. The quadratic function of $v$,$\eta_2 v^2+\eta_1 v+\eta_0$, opens upwards. The symmetric axis satisfies
\bea
v_s=\frac{1}{2}\frac{2C_{6D-18}^p-3C_{5D-15}^p}{C_{6D-18}^p}>\frac{1}{2}~~ \text{for} ~~p\geq7.
\eea
The lower bound of $\eta_2 v^2+\eta_1 v+\eta_0$ is at $v=\frac{1}{2}$,i.e.
\bea
\frac{1}{4}((D-2)(D-4)C_{6D-18}^p+4F_2 C_{4D-12}^p-6(D-2)(D-4)C_{5D-15}^p).
\eea
It is easy to check the above is positive when $p\geq7$. Thus, $S_p>0$ when $p\geq7$.
Then, let's prove
\bea\label{sp-final}
\frac{S_{p+1}}{\bar{f}_{p+1}}>\frac{S_{p}}{\bar{f}_{p}}~\text{for}~p\geq8.
\eea
The above inequality is equivalent to 
\bea\label{sp-final-1}
\frac{S_{p+1}/C_{5D-13}^{p+1}}{\bar{f}_{p+1}/C_{5D-13}^{p+1}}>\frac{S_{p}/C_{5D-13}^{p}}{\bar{f}_{p}/C_{5D-13}^{p}}.
\eea
Consider the following difference
\bea
\frac{S_{p+1}}{C_{5D-13}^{p+1}}-\frac{S_{p}}{C_{5D-13}^{p}}=\frac{(p+1)S_{p+1}-(5D-13-p)S_p}{(5D-13-p)C_{5D-13}^{p}}.
\eea
When $p\geq8$, the numerator of the above difference is positive,i.e.
\bea\label{car-sp}
&&(p+1)S_{p+1}-(5D-13-p)S_p=u^3[(D-5)(D-2)(D-4)C_{6D-18}^pu^3\nn\\
&&+6(D-2)(D-4)C_{5D-15}^p(u+v)u^2-(D-1)F_2(u+v)^2uC_{4D-12}^p\nn\\
&&-2(D-2)(u-v)^2(u+v)C_{3D-9}^p-(3D-7)(D-4)v(G_1 u v + G_2 (u + v)^2)C_{2D-6}^p]\nn\\
&&>0,
\eea
where we use $u>1/2$, $(D-2)(D-4)C_{5D-15}^p(u+v)u^2>2(D-2)(u-v)^2(u+v)C_{3D-9}^p$, $v(G_1 u v + G_2 (u + v)^2)<15u(D-2)/4$, and
\bea
\frac{(D-5)(D-2)(D-4)C_{6D-18}^p/4+5(D-2)(D-4)C_{5D-15}^p/2}{(D-1)F_2C_{4D-12}^p+15(3D-7)(D-4)(D-2)C_{2D-6}^p/4}>1.2~\text{for}~p\geq8.
\eea 
Then, together with Eq.\eqref{bfmc55}, we prove Eq.\eqref{sp-final-1}, i.e. Eq.\eqref{sp-final}. 

With the results in Eqs.\eqref{g1}\eqref{g3-1}\eqref{F1}\eqref{d3}\eqref{sp-final}, we prove that
\bea
\frac{\check{A}^R_{p+1}}{\bar{f}_{p+1}}-\frac{\check{A}^R_p}{\bar{f}_p}>0~~(p\geq8).
\eea
For $1\leq p\leq7$, we can check directly that $\frac{\check{A}^R_{p+1}}{\bar{f}_{p+1}}-\frac{\check{A}^R_p}{\bar{f}_p}>0$ holds.
When $p=1$, the difference $\frac{\check{A}^R_{p+1}}{\bar{f}_{p+1}}-\frac{\check{A}^R_p}{\bar{f}_p}$ is 
\bea
&&2 (D-3)^3 (u-v)^2 ((D-1) u+(D-3) v) ((2 D-5) (5 D-13) u^2\nn\\
&&-(D-1) (5 D-13) u v+(D-2) (D-1) v^2).
\eea
Since $u>v$, it is obviously positive when $D\geq7$. When $p=2$, the difference satisfies
\bea
&&\frac{1}{2} (D - 3)^3 (u - v)^2(k_3 u^3+k_2 u^2v+k_1uv^2+k_0v^3)\nn\\
&&>\frac{1}{2} (D - 3)^3 (u - v)^2((k_3 +k_2+k_1)uv^2+k_0v^3)\nn\\
&&>\frac{1}{2} (D - 3)^3 (u - v)^2(k_3 +k_2+k_1+k_0)v^3>0,
\eea
where
\bea
&&k_3=D (D (2 D (4 D (25 D-281)+4789)-18799)+15881)-3586,\nn\\
&&k_2=(D (D (2 D (D (101 D-1523)+8984)-51829)+73117)-40354),\nn\\
&&k_1=-(D - 2) (D (2 D (4 D^2 + D - 255) + 2143) - 2525),\nn\\
&&k_0=-(D - 3) (D - 2) (2 D (D (5 D - 53) + 175) - 353),
\eea
and we also use the results: $k_3>0,~k_2>0,~k_3+k_2+k_1>0,~k_3+k_2+k_1+k_0>0$ for $D\geq7$.
When $p=3$, the similar argument can be used to prove the difference is positive. When $p=4$,
the difference is 
\bea\label{carp4}
\frac{1}{1440}(D-3)^4(1-2v)(n_4v^4+n_3v^3+n_2v^2+n_1v+n_0),
\eea
 where
\bea
&&n_0=80000 D^8-1445688 D^7+9955851 D^6-27463065 D^5-19290150 D^4\nn\\
&&~~+336996048 D^3-928280156 D^2+1138695960 D-545407200,\nn\\
&&n_1=-307992 D^8+5330579 D^7-33196467 D^6+58112421 D^5+317588127 D^4\nn\\
&&~~-2036657124 D^3+4908514212 D^2-5709234656 D+2657836800,\nn\\
&&n_2=2 (180912 D^8-2705933 D^7+10350693 D^6+49629585 D^5-609429207 D^4\nn\\
&&~~+2467053018 D^3-5154998928 D^2+5589738800 D-2495614080),\nn\\
&&n_3=-8 (16460 D^8-111888 D^7-1939017 D^6+30813387 D^5-187841085 D^4\nn\\
&&~~+621263463 D^3-1176724838 D^2+1203712368 D-517001280),\nn\\
&&n_4=96 (6032 D^7-129296 D^6+1179584 D^5-5934980 D^4+17778503 D^3\nn\\
&&~~-31693094 D^2+31117416 D-12974640).
\eea
Consider the following linear function of $v$
\bea
24n_4 v+6n_3.
\eea
It is easy to check the above function is negative for $0<v<\frac{1}{2}$ when $D\geq7$. Then, the following integral of the above function
\bea
12n_4v^2+6n_3v+2n_2
\eea
is monotonically decreasing in the domain $0<v<\frac{1}{2}$ when $D\geq7$. Its lower bound is at $v=\frac{1}{2}$ and is positive when $D\geq7$. Then, the following integral of the above function
\bea
4n_4v^3+3n_3v^2+2n_2v+n_1
\eea
is monotonically increasing in the domain $0<v<\frac{1}{2}$ when $D\geq7$. Its upper bound is at $v=\frac{1}{2}$ and is negative when $D\geq7$. Then, the following integral of the above function
\bea
n_4v^4+n_3v^3+n_2v^2+n_1v+n_0
\eea
is monotonically decreasing in the domain $0<v<\frac{1}{2}$ when $D\geq7$. Its lower bound is at $v=\frac{1}{2}$ and is positive when $D\geq7$. Thus, it is easy to see that the difference in Eq.\eqref{carp4} is positive, i.e.
\bea
\frac{\check{A}^R_{5}}{\bar{f}_{5}}-\frac{\check{A}^R_4}{\bar{f}_4}>0.
\eea
We can use the same strategy to prove differences of the $p=5,6,7$ cases are also positive. Thus, we finish the proof of Eq.\eqref{cAR-final}.

\subsection{Proof of $\frac{\bar{B}^M_{p+1}}{\bar{f}_{p+1}}>\frac{\bar{B}^M_p}{\bar{f}_p}$}
\label{sec-bbm}
We first prove 
\bea\label{bBmp-positive}
\bar{B}^M_p>0~~(p\geq1).
\eea
Since $2u^2(D-3)$ is positive, we just consider the rest factor of $\bar{B}^M_p$, i.e.,
\bea\label{bBmp-factor}
C_{5D-13}^p u (u + v) - C_{4D-10}^p (u^2 + 4 u v + v^2)+3 C_{3D-7}^p (u + v)v-2 C_{2D-4}^p v^2.
\eea
Replacing $u$ with $(1-v)$, we obtain
\bea\label{bBmp-factor-v}
2v^2 (C_{4D-10}^p-C_{2D-4}^p)+v (3C_{3D-7}^p-2C_{4D-10}^p-C_{5D-13}^p)+C_{5D-13}^p-C_{4D-10}^p.
\eea
The above quadratic function of $v$ has a positive intersection and opens upwards. Its symmetric axis satisfies
\bea
v_s=\frac{1}{2}\frac{C_{5D-13}^p+2C_{4D-10}^p-3C_{3D-7}^p}{2(C_{4D-10}^p-C_{2D-4}^p)}\geq\frac{1}{2}.
\eea
Proof of the above inequality. It is easy to see that $C_{5D-13}^p+2C_{4D-10}^p-3C_{3D-7}^p>0$ and $2(C_{4D-10}^p-C_{2D-4}^p>0$.
Then, their difference is 
\bea
&&(C_{5D-13}^p+2C_{4D-10}^p-3C_{3D-7}^p)-(2(C_{4D-10}^p-C_{2D-4}^p)\nn\\
&&=C_{5D-13}^p+2C_{2D-4}^p-3C_{3D-7}^p.
\eea
When $p\geq3$, $\frac{C_{5D-13}^p}{C_{3D-7}^p}>3$, the above expression is positive. When $p=1$, the above expression is zero. When $p=2$, the above expression equals to
$3(D-3)^2$, which is positive. \\
Since $0<v<1/2$, the lower bound of Eq.\eqref{bBmp-factor-v} is at $v=1/2$, which is
\bea
\frac{1}{2}(C_{5D-13}^p-3C_{4D-10}^p+3C_{3D-7}^p-C_{2D-4}^p).
\eea 
When $p=1,2,3,4,5$, the values of the above lower bound are respectively $\{0,0,(D-3)^3, (D-3)^3 (7 D-20)/2, (D-3)^3 (5 D-16) (5 D-14)/4\}$, which are all positive.
When $p\geq6$, $\frac{C_{5D-13}^p}{C_{4D-10}^p}>3$ and the above lower bound is also positive. So we prove Eq.\eqref{bBmp-positive}.

Now we prove
\bea\label{bBmu-final}
\frac{\bar{B}^M_{p+1}}{\bar{f}_{p+1}}-\frac{\bar{B}^M_p}{\bar{f}_p}>0~~(p\geq1).
\eea
For $p=1,2,3$, we calculate the differences directly and prove they are positive. When $p=1$, we have
\bea
&&\frac{\bar{B}^M_{2}}{\bar{f}_{2}}-\frac{\bar{B}^M_1}{\bar{f}_1}\nn\\
&=&\frac{2 (D - 3) u (u - v)}{\bar{f}_2\bar{f}_1}
((10 D^2-51 D+65) u^2+(-5 D^2+18 D-13) u v+(D^2-3 D+2) v^2)\nn\\
&>&\frac{2 (D - 3) u (u - v)}{\bar{f}_2\bar{f}_1}((5 D^2-33 D+52)uv+(D^2-3 D+2) v^2)>0,
\eea
where we use the facts that $u>v>0$ and $10 D^2-51 D+65>0,5 D^2-33 D+52>0$ for $D\geq7$. When $p=2$, we have
\bea
&&\frac{\bar{B}^M_{3}}{\bar{f}_{3}}-\frac{\bar{B}^M_2}{\bar{f}_2}\nn\\
&=&\frac{(D - 3) u}{3\bar{f}_3\bar{f}_2}(\g_0 u^3+\g_1 u^2v+\g_2u v^2+\g_3v^3)
>\frac{(D - 3) u}{3\bar{f}_3\bar{f}_2}((\g_0 +\g_1) u^2v+\g_2u v^2+\g_3v^3)\nn\\
&>&\frac{(D - 3) u}{3\bar{f}_3\bar{f}_2}((\g_0 +\g_1+\g_2)u v^2+\g_3v^3)>0,
\eea
where
\bea
&&\g_0=200 D^4 - 2130 D^3 + 8501 D^2 - 15069 D + 10010, \nn\\
&&\g_1=-3 (25 D^4 - 210 D^3 + 612 D^2 - 681 D + 182),\nn\\
&&\g_2=-3 (6 D^4 - 84 D^3 + 401 D^2 - 801 D + 574),\nn\\
&&\g_3=D^4 - 12 D^3 +46 D^2 - 69 D + 34,
\eea
and we also use the facts that $u>v>0$ and $\g_0>0,~\g_0+\g_1>0,~\g_0+\g_1+\g_2>0,~\g_3>0$ for $D\geq7$. The proof of $p=3$ case is similar to the proof of $p=2$ case.
When $p\geq4$, we have
\bea
&&\frac{\bar{B}^M_{p+1}}{\bar{f}_{p+1}}-\frac{\bar{B}^M_p}{\bar{f}_p}\nn\\
&=&\frac{2(D-3)u}{(p+1)\bar{f}_{p+1}\bar{f}_{p}}\left(-6 C_{5D-13}^p v (u + v) (-C_{2D-4}^p v + C_{3D-7}^p (u + v))\right.\nn\\
&&\left. +C_{4D-10}^p (-8 C_{2D-4}^p v^3 + 6 C_{3D-7}^p v^2 (u + v) + C_{5D-13}^p (u + v)^3)\right).
\eea 
Replacing $u$ with $1-v$ in the above equation, and since $v<1/2$, we obtain
\bea\label{bBmu-pgeq4}
&&\frac{\bar{B}^M_{p+1}}{\bar{f}_{p+1}}-\frac{\bar{B}^M_p}{\bar{f}_p}\nn\\
&=&\frac{2(D-3)u}{(p+1)\bar{f}_{p+1}\bar{f}_{p}}\left(C_{4D-10}^p C_{5D-13}^p - 6 C_{3D-7}^p C_{5D-13}^p v + 6 (C_{3D-7}^p C_{4D-10}^p + C_{2D-4}^p C_{5D-13}^p) v^2 \right.\nn\\
&&\left.- 8 C_{2D-4}^p C_{4D-10}^p v^3\right)\nn\\
&>&\frac{2(D-3)u}{(p+1)\bar{f}_{p+1}\bar{f}_{p}}\left(C_{4D-10}^p C_{5D-13}^p - 6 C_{3D-7}^p C_{5D-13}^p v + 6 (C_{3D-7}^p C_{4D-10}^p + C_{2D-4}^p C_{5D-13}^p) v^2 \right.\nn\\
&&\left.- 4 C_{2D-4}^p C_{4D-10}^p v^2\right).
\eea
Since $\frac{2(D-3)u}{(p+1)\bar{f}_{p+1}\bar{f}_{p}}>0$ we just consider the rest factor, i.e.,
\bea
C_{4D-10}^p C_{5D-13}^p - 6 C_{3D-7}^p C_{5D-13}^p v + 6 (C_{3D-7}^p C_{4D-10}^p + C_{2D-4}^p C_{5D-13}^p) v^2
- 4 C_{2D-4}^p C_{4D-10}^p v^2.\nn
\eea
Taking it as a quadratic function of $v(0<v<\frac{1}{2})$, it is obvious that this function has a positive intersection and opens upwards. The symmetric axis is at
\bea
v_s=\frac{1}{2}\cdot\frac{6 C_{3D-7}^p C_{5D-13}^p}{6 (C_{3D-7}^p C_{4D-10}^p + C_{2D-4}^p C_{5D-13}^p)- 4 C_{2D-4}^p C_{4D-10}^p}.
\eea
Since $1-\frac{C_{4D-10}^p}{C_{5D-13}^p}-\frac{C_{2D-4}^p}{C_{3D-7}^p}>0.3$ for $p\geq4$ we have $v_s>\frac{1}{2}$. The lower bound of the quadratic function is at $v=\frac{1}{2}$, which is 
\bea
C_{4D-10}^p(\frac{3}{2}C_{3D-7}^p-C_{2D-4}^p)+\frac{3}{2}C_{2D-4}^p C_{5D-13}^p+C_{5D-13}^p(C_{4D-10}^p-3C_{3D-7}^p).\nn
\eea
Because $\frac{C_{4D-10}^p}{C_{3D-7}^p}>3$ for $p\geq4$, the above lower bound is positive. Then, Eq.\eqref{bBmu-pgeq4} is positive. We finish the proof of Eq.\eqref{bBmu-final}.

\subsection{Proof of $\frac{\bar{B}^R_{p+1}}{\bar{f}_{p+1}}>\frac{\bar{B}^R_p}{\bar{f}_p}$}
\label{sec-bbr}
Now we prove
\bea\label{bBR-final}
\frac{\bar{B}^R_{p+1}}{\bar{f}_{p+1}}-\frac{\bar{B}^R_p}{\bar{f}_p}>0~~(p\geq1).
\eea
An important observation is that the difference $\frac{\bar{B}^R_{p+1}}{\bar{f}_{p+1}}-\frac{\bar{B}^R_p}{\bar{f}_p}$ is linear in $\o$. In order to prove the positivity of the 
difference, we just need to check the $\o=0$ and $\o=\frac{c_D eq}{u}$ cases (remember the superradiance condition $0<\o<\frac{c_D eq}{u}$). When $\o=0$,
the difference is 
\bea
\left(\frac{\bar{B}^R_{p+1}}{\bar{f}_{p+1}}-\frac{\bar{B}^R_p}{\bar{f}_p}\right)_{\o=0}=\frac{4(D-3)^2u(u+v)c_D^2e^2q^2}{\bar{f}_{p+1}\bar{f}_p}
(\d_0 u^2+\d_1 uv+\d_2v^2)
\eea
where
\bea\label{delta}
&&\d_0=C_{4D-10}^p C_{5D-13}^p,\nn\\
&&\d_1=2(C_{3D-7}^p- C_{2D-4}^p) C_{4D-10}^p +  C_{3D-7}^p C_{4D-10}^p - 3 C_{2D-4}^p C_{5D-13}^p,\nn\\
&&\d_2=C_{2D-4}^p (3 C_{3D-7}^p - 2 C_{4D-10}^p).
\eea
For $p=1$, 
\bea
\d_0=2(2D-5)(5D-13),\d_1=-2(D-1)(5D-13),\d_2=2(D-1)(D-2).
\eea
Since $u>v>0$ we have $\d_0 u^2+\d_1 uv+\d_2v^2>(\d_0+\d_1) uv+\d_2v^2>0$. For $p\geq2$,
we have 
\bea
\d_0 u^2+\d_1 uv+\d_2v^2>(\d_0+\d_1) uv+\d_2v^2>(\d_0+\d_1+\d_2)v^2>0,
\eea
where we use the facts that $\frac{C_{4D-10}^p}{C_{2D-4}^p}>3$ and $\frac{3C_{3D-7}^p}{4C_{2D-4}^p}>1.5$. So we obtain
\bea
\left(\frac{\bar{B}^R_{p+1}}{\bar{f}_{p+1}}-\frac{\bar{B}^R_p}{\bar{f}_p}\right)_{\o=0}>0.
\eea
When $\o=\frac{c_D e q}{u}$,
we find that
\bea
\left(\frac{\bar{B}^R_{p+1}}{\bar{f}_{p+1}}-\frac{\bar{B}^R_p}{\bar{f}_p}\right)_{\o=\frac{c_D e q}{u}}=\frac{4(D-3)^2u(u-v)c_D^2e^2q^2}{\bar{f}_{p+1}\bar{f}_p}
(\d_0 u^2+\d_1 uv+\d_2v^2)>0.
\eea
Then we finish the proof of Eq.\eqref{bBR-final}.

\section{$\text{sign}(b_{p+1})\geq \text{sign}(b_{p})$, $ D-3\leq p\leq 2D-7$}

When $ D-2\leq p\leq 2D-6$, the coefficient of $z^p$ is 
\bea\label{b-2d6d2}
b_p&=&a_6 C_{6D-18}^{p}r_h^{6D-18-p}+ a_5 C_{5D-15}^{p}r_h^{5D-15-p}\nn\\
&&+a_4 C_{4D-12}^{p}r_h^{4D-12-p}+a_3 C_{3D-9}^{p}r_h^{3D-9-p}+a_2 C_{2D-6}^{p}r_h^{2D-6-p}\nn\\
&&+a'_5 C_{5D-13}^{p}r_h^{5D-13-p}+a'_4 C_{4D-10}^{p}r_h^{4D-10-p}+a'_3C_{3D-7}^{p}r_h^{3D-7-p}+a'_2C_{2D-4}^{p}r_h^{2D-4-p}\nn\\
&=&r_h^{-p}(a_6 u^6 C_{6D-18}^{p}+ a_5 u^5 C_{5D-15}^{p}+a_4u^4 C_{4D-12}^{p}+a_3 u^3C_{3D-9}^{p}+a_2 u^2 C_{2D-6}^{p})\nn\\
&&+r_h^{-p+2}u^2(a'_5u^3 C_{5D-13}^{p}+a'_4 u^2 C_{4D-10}^{p}+a'_3 u C_{3D-7}^{p}+a'_2C_{2D-4}^{p}).
\eea
Define the normalized coefficient as
\bea
b'_p=\frac{r_h^p}{\bar{f}_p}b_p=\frac{\tilde{A}_p}{\bar{f}_p}+r_h^2u^2\frac{\bar{B}_p}{\bar{f}_p}
=\frac{\tilde{A}^L_p}{\bar{f}_p}\l_l+\frac{\tilde{A}^R_p}{\bar{f}_p}+r_h^2u^2\frac{\bar{B}_p}{\bar{f}_p},
\eea
where 
\bea
&&\tilde{A}^{L}_p=2u^4\left(2u^2 C_{6D-18}^{p}+(D-7)u(u+v) C_{5D-15}^{p}-(D-5)(u^2+4uv+v^2) C_{4D-12}^{p}\right.\nn\\\nn
&&~~~~ \left.+(3D-13)v(u+v) C_{3D-9}^{p}-(2D-8)v^2 C_{2D-6}^{p} \right)\\\nn
&&\tilde{A}^{R}_p=u^3\left((D-2)(D-4)u^3C_{6D-18}^p-3(D-2)(D-4)u^2(u+v)C_{5D-15}^{p} \right.\\\nn
&&~~~~\left.+u(( 4 D^3- 29 D^2 + 66 D-48)uv-(D^3-11 D^2+ 39 D-42)(u+v)^2)C_{4D-12}^{p}\right.\\\nn
&&~~~~\left.+(u + v)((u + v)^2-2(5D^2-30D+42) u v)C_{3D-9}^{p}\right.\\
&&~~~~\left.+(D-4)v((D^2-3D+3)(u + v)^2-(4D^2-27D+42)uv)C_{2D-6}^{p}\right).
\eea
Since $\tilde{A}^L_p=\check{A}^L_p$, according to the proof in Sec.\eqref{sec-cal}, we have
\bea
\frac{\tilde{A}^L_{p+1}}{\bar{f}_{p+1}}>\frac{\tilde{A}^L_p}{\bar{f}_p}.
\eea
According to the proofs in Sec.\eqref{sec-bbm} and Sec.\eqref{sec-bbr}, we also have
\bea
r_h^2u^2\frac{\bar{B}_{p+1}}{\bar{f}_{p+1}}>r_h^2u^2\frac{\bar{B}_p}{\bar{f}_p}.
\eea
The difference between $\tilde{A}^R_p$ and $\check{A}^R_p$ is the $g_1$ term in Eq.\eqref{car-gi}. In the proof in Sec.\eqref{sec-car}, the $g_1$ term and other terms 
are discussed separately, so we have
\bea
\frac{\tilde{A}^R_{p+1}}{\bar{f}_{p+1}}>\frac{\tilde{A}^R_p}{\bar{f}_p}.
\eea

Thus, we prove
\bea\label{b'-d-2-2d-5}
b'_{p+1}>b'_p,(D-2\leq p\leq 2D-5),
\eea 
and obtain
\bea\label{sign-d-2-2d-5}
\text{sign}(b_{p+1})\geq \text{sign}(b_{p}), (D-2\leq p\leq 2D-5).
\eea

When $p=D-3$, we need prove $ b'_{D-2}> b'_{D-3}$, i.e.
\bea
\frac{\tilde{A}_{D-2}}{\bar{f}_{D-2}}+r_h^2u^2\frac{\bar{B}_{D-2}}{\bar{f}_{D-2}}>\frac{\check{A}_{D-3}}{\bar{f}_{D-3}}+r_h^2u^2\frac{\bar{B}_{D-3}}{\bar{f}_{D-3}}.
\eea
Based on the proof of Eq.\eqref{b'-d-2-2d-5}, it is easy to deduce
\bea
\frac{\tilde{A}_{D-2}}{\bar{f}_{D-2}}+r_h^2u^2\frac{\bar{B}_{D-2}}{\bar{f}_{D-2}}>\frac{\tilde{A}_{D-3}}{\bar{f}_{D-3}}+r_h^2u^2\frac{\bar{B}_{D-3}}{\bar{f}_{D-3}}.
\eea
According to the definitions of $\check{A}_p$ and  $\tilde{A}_p$, we have
\bea
\check{A}_{D-3}=\tilde{A}_{D-3}-3(D-2)(D-4)u^3(u+v)v^2C_{D-3}^{p}<\tilde{A}_{D-3},
\eea
and then
\bea\nn
\frac{\tilde{A}_{D-2}}{\bar{f}_{D-2}}+r_h^2u^2\frac{\bar{B}_{D-2}}{\bar{f}_{D-2}}>\frac{\check{A}_{D-3}}{\bar{f}_{D-3}}+r_h^2u^2\frac{\bar{B}_{D-3}}{\bar{f}_{D-3}}.
\eea
Thus, we have
\bea\label{sign-d-2-d-3}
\text{sign}(b_{p+1})\geq \text{sign}(b_{p}),~(p=D-3).
\eea

\section{$\text{sign}(b_{p+1})\geq \text{sign}(b_{p})$, $p= 2D-6,2D-5$}

The coefficient $b_p$ ($p=2D-4, 2D-5$) is 
\bea\label{b-2d42d5}
b_p&=&a_6 C_{6D-18}^{p}r_h^{6D-18-p}+ a_5 C_{5D-15}^{p}r_h^{5D-15-p}+a_4 C_{4D-12}^{p}r_h^{4D-12-p}+a_3 C_{3D-9}^{p}r_h^{3D-9-p}\nn\\
&&+a'_5 C_{5D-13}^{p}r_h^{5D-13-p}+a'_4 C_{4D-10}^{p}r_h^{4D-10-p}+a'_3C_{3D-7}^{p}r_h^{3D-7-p}+a'_2C_{2D-4}^{p}r_h^{2D-4-p}\nn\\
&=&r_h^{-p}(a_6 u^6 C_{6D-18}^{p}+ a_5 u^5 C_{5D-15}^{p}+a_4u^4 C_{4D-12}^{p}+a_3 u^3C_{3D-9}^{p})\nn\\
&&+r_h^{-p+2}u^2(a'_5u^3 C_{5D-13}^{p}+a'_4 u^2 C_{4D-10}^{p}+a'_3 u C_{3D-7}^{p}+a'_2C_{2D-4}^{p}).
\eea
Define the normalized coefficient as
\bea
b'_{p}=\frac{r_h^p}{\bar{f}_p}b_p=\frac{\bar{A}_p}{\bar{f}_p}+r_h^2u^2\frac{\bar{B}_p}{\bar{f}_p}=\frac{\bar{A}^L_p}{\bar{f}_p}\l_l+\frac{\bar{A}^R_p}{\bar{f}_p}
+r_h^2u^2\frac{\bar{B}_p}{\bar{f}_p},
\eea
where
\bea\label{bALp-bARp}\nn
&&\bar{A}^{L}_p=2u^4\left(2u^2 C_{6D-18}^{p}+(D-7)u(u+v) C_{5D-15}^{p}-(D-5)(u^2+4uv+v^2) C_{4D-12}^{p}\right.\\\nn
&&~~~~ \left.+(3D-13)v(u+v) C_{3D-9}^{p}\right),\\\nn
&&\bar{A}^{R}_p=u^3\left((D-2)(D-4)u^3C_{6D-18}^p-3(D-2)(D-4)u^2(u+v)C_{5D-15}^{p} \right.\\\nn
&&~~~~\left.+u(( 4 D^3- 29 D^2 + 66 D-48)uv-(D^3-11 D^2+ 39 D-42)(u+v)^2)C_{4D-12}^{p}\right.\\
&&~~~~\left.+(u + v)((u + v)^2-2(5D^2-30D+42) u v)C_{3D-9}^{p}\right).
\eea

Now, let's prove 
\bea
&&b'_{2D-5}=\frac{\bar{A}^L_{2D-5}}{\bar{f}_{2D-5}}\l_l+\frac{\bar{A}^R_{2D-5}}{\bar{f}_{2D-5}}
+r_h^2u^2\frac{\bar{B}_{2D-5}}{\bar{f}_{2D-5}}\nn\\
&>&b'_{2D-6}=\frac{\tilde{A}^L_{2D-6}}{\bar{f}_{2D-6}}\l_l+\frac{\tilde{A}^R_{2D-6}}{\bar{f}_{2D-6}}
+r_h^2u^2\frac{\bar{B}_{2D-6}}{\bar{f}_{2D-6}}.
\eea
Since $2D-6>7$ for $D\geq7$, according to the proof in Sec.\eqref{sec-cal}, we have 
\bea
\frac{\bar{f}_{2D-5}}{C_{5D-15}^{2D-5}}<\frac{\bar{f}_{2D-6}}{C_{5D-15}^{2D-6}},
\eea
and 
\bea
\frac{\bar{A}^L_{2D-5}}{C_{5D-15}^{2D-5}}-\frac{\tilde{A}^L_{2D-6}}{C_{5D-15}^{2D-6}}
>\frac{2u^4}{(3D-7)C_{5D-13}^{2D-6}}((\b_0+\b_1)u v+\b_2 v^2)>0,
\eea
where $\beta_0,\beta_1,\b_2$ are defined in Eq.\eqref{beta012} with $p=2D-6$. Thus,
we get
\bea
\frac{\bar{A}^L_{2D-5}}{\bar{f}_{2D-5}}\l_l>\frac{\tilde{A}^L_{2D-6}}{\bar{f}_{2D-6}}\l_l.
\eea
Similarly, according to the proof in Sec.\eqref{sec-car}, we can also prove 
\bea
\frac{\bar{A}^R_{2D-5}}{\bar{f}_{2D-5}}>\frac{\tilde{A}^R_{2D-6}}{\bar{f}_{2D-6}}.
\eea
Two points should be emphasized here. One is that $g_1$ term and other terms separately satisfy the required inequality in the proof in Sec.\eqref{sec-car}.
The second is that in the proof in Sec.\eqref{sec-car}, all the $C_\ast^{p+1}$ is rewritten as $\frac{\ast-p}{p+1}C_\ast^p$, and then formally we have 
\bea
\bar{A}^R_{2D-5}=\tilde{A}^R_{2D-5}
\eea
since 
\bea
\bar{A}^R_{2D-5}=\tilde{A}^R_{2D-5}-u^3g_2C_{2D-6}^{2D-5}=\tilde{A}^R_{2D-5}-u^3g_2\frac{0}{2D-5}=\tilde{A}^R_{2D-5}.
\eea
According to the proofs in Sec.\eqref{sec-bbm} and \eqref{sec-bbr}, we have
\bea
r_h^2u^2\frac{\bar{B}_{2D-5}}{\bar{f}_{2D-5}}>r_h^2u^2\frac{\bar{B}_{2D-6}}{\bar{f}_{2D-6}}.
\eea

Thus, we have $ b'_{2D-5}> b'_{2D-6}$ and
\bea\label{sign-2d-5-2d-6}
\text{sign}(b_{p+1})\geq \text{sign}(b_{p}),(p=2D-6).
\eea

Then, let's prove 
\bea
&&b'_{2D-4}=\frac{\bar{A}^L_{2D-4}}{\bar{f}_{2D-4}}\l_l+\frac{\bar{A}^R_{2D-4}}{\bar{f}_{2D-4}}
+r_h^2u^2\frac{\bar{B}_{2D-4}}{\bar{f}_{2D-4}}\nn\\
&>&b'_{2D-5}=\frac{\bar{A}^L_{2D-5}}{\bar{f}_{2D-5}}\l_l+\frac{\bar{A}^R_{2D-5}}{\bar{f}_{2D-5}}
+r_h^2u^2\frac{\bar{B}_{2D-5}}{\bar{f}_{2D-5}}.
\eea
According to the proofs in Sec.\eqref{sec-bbm} and \eqref{sec-bbr}, we have
\bea
r_h^2u^2\frac{\bar{B}_{2D-4}}{\bar{f}_{2D-4}}>r_h^2u^2\frac{\bar{B}_{2D-5}}{\bar{f}_{2D-5}}.
\eea
By taking the $C_{2D-6}^p$ term to zero in the proof of Eq.\eqref{cALmc55} in Sec.\eqref{sec-cal}, it is easy to get
\bea
\frac{\bar{A}^L_{2D-4}}{\bar{f}_{2D-4}}>\frac{\bar{A}^L_{2D-5}}{\bar{f}_{2D-5}}.
\eea
Then, for the term including $\bar{A}^R_{p}$, we have  
\bea\label{bar-gi}
\bar{A}^R_{p}=u^3\sum_{i=3}^{6}g_i C_{i(D-3)}^p,(p=2D-5,2D-4),
\eea
where
\bea
&&g_6=(D-2)(D-4)u^3,g_5=-3(D-2)(D-4)u^2(u+v),\nn\\
&&g_4=-(D-2)^3u(u-v)^2+F_2(u+v)^2u-F_1u^2v,\nn\\
&&g_3=(u-v)^2(u+v)-10(D-2)(D-4)(u+v)uv,
\eea
and
\bea
F_1=5D^2-18D+16,~F_2=5D^2-27D+34.
\eea
The proofs for the negative terms in $g_3,g_4$ are the same as that in Eqs.\eqref{g3-1}\eqref{F1}. 
The proof for other terms is the same as that for Eqs.\eqref{sp-final}\eqref{sp-final-1}. 
It is easy to check that after taking the $C_{2D-6}^p$ term to zero, all the inequalities in the proof of Eqs.\eqref{sp-final}\eqref{sp-final-1} still hold. So, we obtain
\bea\label{bar2d-4}
\frac{\bar{A}^R_{2D-4}}{\bar{f}_{2D-4}}>\frac{\bar{A}^R_{2D-5}}{\bar{f}_{2D-5}}.
\eea

Thus, we have $ b'_{2D-4}> b'_{2D-5}$ and
\bea\label{sign-2d-4-2d-5}
\text{sign}(b_{p+1})\geq \text{sign}(b_{p}),(p=2D-5).
\eea

\section{$\text{sign}(b_{p+1})\geq \text{sign}(b_{p})$, $ 2D-4\leq p\leq 3D-10$}
\label{sec-2d-4-3d-10}
For $ 2D-3\leq p\leq 3D-9$, The coefficient of $b_p$ is 
\bea
b_p&=&a_6 C_{6D-18}^{p}r_h^{6D-18-p}+ a_5 C_{5D-15}^{p}r_h^{5D-15-p}+a_4 C_{4D-12}^{p}r_h^{4D-12-p}+a_3 C_{3D-9}^{p}r_h^{3D-9-p}\nn\\
&&+a'_5 C_{5D-13}^{p}r_h^{5D-13-p}+a'_4 C_{4D-10}^{p}r_h^{4D-10-p}+a'_3 C_{3D-7}^{p}r_h^{3D-7-p}\nn\\
&=& r_h^{-p}(a_6 u^6 C_{6D-18}^{p}+ a_5 u^5 C_{5D-15}^{p}+a_4 u^4 C_{4D-12}^{p}+a_3 u^3 C_{3D-9}^{p} )\nn\\
&&+r_h^{-p+2}u^2(a'_5 u^3 C_{5D-13}^{p}+a'_4 u^2 C_{4D-10}^{p}+a'_3 u C_{3D-7}^{p})
\eea
The normalized coefficient is defined as
\bea
b'_{p}=\frac{r_h^{p}}{\hat{f}_{p}}b_{p}=\frac{\bar{A}_p}{\hat{f}_{p}}+r_h^2u^2\frac{\hat{B}_p}{\hat{f}_{p}}
=\frac{\bar{A}^L_p}{\hat{f}_p}\l_l+\frac{\bar{A}^R_p}{\hat{f}_p}+r_h^2u^2(\frac{\hat{B}^M_p}{\hat{f}_p}(\mu^2-\o^2)+\frac{\hat{B}^R_p}{\hat{f}_p}),
\eea
where
\bea\label{hafp-hatbp}
&&\hat{f}_p=u(u+v)C_{5D-13}^{p}+(u^2+v^2)C_{4D-10}^{p}-3v(u+v)C_{3D-7}^{p},\nn\\
&&\hat{B}_p=a'_5 u^3 C_{5D-13}^{p}+a'_4 u^2 C_{4D-10}^{p}+a'_3 u C_{3D-7}^{p},\nn\\
&&\hat{B}^M_p=2 u^2(D-3)\left(C_{5D-13}^p u (u + v) - C_{4D-10}^p (u^2 + 4 u v + v^2)
+3 C_{3D-7}^p (u + v)v\right),\nn\\
&&\hat{B}^R_p=2 (D-3) u \left(-u\hat{f}_p \o^2+2c_D e q u(C_{5D-13}^p u+C_{4D-10}^p(u+v)-3C_{3D-7}^p v)\o\right.\nn\\
&&~~~~~\left.-2c_D^2 e^2 q^2 u C_{4D-10}^p\right),
\eea
and $\bar{A}^L_p,\bar{A}^R_p$ are defined in Eq.\eqref{bALp-bARp}.

According to the proof of Eq.\eqref{bf-positive}, it is easy to know that $\hat{f}_p>0$. 
According to the proof of Eq.\eqref{bBmp-positive}, it is easy to know that $\hat{B}^M_p>0$. 
Based on the proof in Eq.\eqref{bBmu-pgeq4} and taking the $C_{2D-4}^p$ term to zero,  we have
\bea\label{hBmu}
&&\frac{\hat{B}^M_{p+1}}{\hat{f}_{p+1}}-\frac{\hat{B}^M_p}{\hat{f}_p}\\
&=&\frac{2(D-3)u}{(p+1)\hat{f}_{p+1}\hat{f}_{p}}\left(C_{4D-10}^p C_{5D-13}^p - 6v C_{3D-7}^p C_{5D-13}^p  + 6 C_{3D-7}^p C_{4D-10}^p v^2\right)>0,\nn 
\eea
where we use $\frac{C_{4D-10}^p}{C_{3D-7}^p}>17$ when $p\geq 2D-3\geq11$. Based on the proof in Sec.\eqref{sec-bbr}, it is easy to deduce 
\bea\label{hBr}
\frac{\hat{B}^R_{p+1}}{\hat{f}_{p+1}}-\frac{\hat{B}^R_p}{\hat{f}_p}>0,
\eea
since $\d_0,\d_1,\d_2$ defined in Eq.\eqref{delta} are all positive in this case.

Following the proof in Eq.\eqref{bfmc55}, it is easy to deduce
\bea
\frac{\hat{f}_{p}}{C_{5D-13}^{p}}>\frac{\hat{f}_{p+1}}{C_{5D-13}^{p+1}}.
\eea
Following the proof in Eq.\eqref{cALmc55}, it is easy to deduce
\bea
\frac{\bar{A}^L_{p+1}}{C_{5D-13}^{p+1}}>\frac{\bar{A}^L_p}{C_{5D-13}^{p}},
\eea
since $\b_0,\b_1$ defined in Eq.\eqref{beta012} preserve the same form and $\b_2$ is still positive.  
So, we have
\bea\label{balp-final}
\frac{\bar{A}^L_{p+1}}{\hat{f}_{p+1}}>\frac{\bar{A}^L_p}{\hat{f}_p}.
\eea
Following the proof in Sec.\eqref{sec-car}, and taking terms proportional to $C_{D-3}^p,C_{2D-6}^p$ and $y_2$ as zero, it is easy to deduce
\bea
\frac{\bar{A}^R_{p+1}}{\hat{f}_{p+1}}>\frac{\bar{A}^R_p}{\hat{f}_p}.
\eea

Thus, we have $ b'_{p+1}> b'_{p}$ and
\bea\label{sign-2d-3-3d-10}
\text{sign}(b_{p+1})\geq \text{sign}(b_{p}),~(2D-3\leq p\leq 3D-10).
\eea

Now, let's prove $b'_{2D-3}>b'_{2D-4}$, i.e.
\bea
&&\frac{\bar{A}^L_{2D-3}}{\hat{f}_{2D-3}}\l_l+\frac{\bar{A}^R_{2D-3}}{\hat{f}_{2D-3}}+r_h^2u^2(\frac{\hat{B}^M_{2D-3}}{\hat{f}_{2D-3}}(\mu^2-\o^2)+\frac{\hat{B}^R_{2D-3}}{\hat{f}_{2D-3}})\nn\\
&>&\frac{\bar{A}^L_{2D-4}}{\bar{f}_{2D-4}}\l_l+\frac{\bar{A}^R_{2D-4}}{\bar{f}_{2D-4}}+r_h^2u^2(\frac{\bar{B}^M_{2D-4}}{\bar{f}_{2D-4}}(\mu^2-\o^2)+\frac{\bar{B}^R_{2D-4}}{\bar{f}_{2D-4}}).
\eea
Since $\bar{f}_{2D-4}>\hat{f}_{2D-4}>0$ and $\bar{A}^L_{2D-4}>0$, we have
\bea
\frac{\bar{A}^L_{2D-4}}{\bar{f}_{2D-4}}<\frac{\bar{A}^L_{2D-4}}{\hat{f}_{2D-4}}<\frac{\bar{A}^L_{2D-3}}{\hat{f}_{2D-3}},
\eea 
where we use the result in Eq.\eqref{balp-final} in the last inequality. 
Similarly, since $\bar{f}_{2D-4}>\hat{f}_{2D-4}>0$ and $\hat{B}^M_{2D-4}>\bar{B}^M_{2D-4}>0$, we have
\bea
\frac{\bar{B}^M_{2D-4}}{\bar{f}_{2D-4}}<\frac{\hat{B}^M_{2D-4}}{\hat{f}_{2D-4}}<\frac{\hat{B}^M_{2D-3}}{\hat{f}_{2D-3}}.
\eea

The difference $\frac{\hat{B}^R_{2D-3}}{\hat{f}_{2D-3}}-\frac{\bar{B}^R_{2D-4}}{\bar{f}_{2D-4}}$ is linear in $\omega$. For $\o=0$,
\bea
&&\left(\frac{\hat{B}^R_{2D-3}}{\hat{f}_{2D-3}}-\frac{\bar{B}^R_{2D-4}}{\bar{f}_{2D-4}}\right)_{\o=0}=\frac{4c_D^2 e^2 q^2(D-3)u}{\hat{f}_{2D-3}\bar{f}_{2D-4}}
(\hat{f}_{2D-3}(C_{4D-10}^{2D-4}u-v)-\bar{f}_{2D-4}C_{4D-10}^{2D-3}u)\nn\\
&&=\frac{4c_D^2 e^2 q^2(D-3)^2u}{\hat{f}_{2D-3}\bar{f}_{2D-4}(2D-3)}
(\epsilon_0 u^3+\epsilon_1u^2v+\epsilon_2u v^2+\epsilon_3v^3)\nn\\
&&>\frac{4c_D^2 e^2 q^2(D-3)^2u}{\hat{f}_{2D-3}\bar{f}_{2D-4}(2D-3)}
((\epsilon_0 +\epsilon_1+\epsilon_2)u v^2+\epsilon_3v^3)\nn\\
&&>\frac{4c_D^2 e^2 q^2(D-3)^2u}{\hat{f}_{2D-3}\bar{f}_{2D-4}(2D-3)}
(\epsilon_0 +\epsilon_1+\epsilon_2+\epsilon_3)v^3>0,
\eea
where
\bea
&&\epsilon_0=C_{4D-10}^{2D-4}C_{5D-13}^{2D-4},\nn\\
&&\epsilon_1=-2 C_{4D-10}^{2D-4} + 3 C_{3D-7}^{2D-4} C_{4D-10}^{2D-4} - 3 C_{5D-13}^{2D-4} + C_{4D-10}^{2D-4}C_{5D-13}^{2D-4},\nn\\
&&\epsilon_2=3C_{3D-7}^{2D-4} - 4 C_{4D-10}^{2D-4} + 3C_{3D-7}^{2D-4} C_{4D-10}^{2D-4} - 3C_{5D-13}^{2D-4},\nn\\
&&\epsilon_3= 3 C_{3D-7}^{2D-4} - 2 C_{4D-10}^{2D-4},
\eea
and we also use the inequalities $\ep_0>0,\ep_1>0,(\ep_0+\ep_1+\ep_2)>0,(\ep_0+\ep_1+\ep_2+\ep_3)>0$, which are easy to check.
For $\o=\frac{c_D eq}{u}$,
\bea
&&\left(\frac{\hat{B}^R_{2D-3}}{\hat{f}_{2D-3}}-\frac{\bar{B}^R_{2D-4}}{\bar{f}_{2D-4}}\right)_{\o=\frac{c_D eq}{u}}
=\frac{4c_D^2 e^2 q^2(D-3)^2u(u-v)}{\hat{f}_{2D-3}\bar{f}_{2D-4}(2D-3)}
(\zeta_0 u^2+\zeta_1u v+\zeta_2v^2)\nn\\
&&>\frac{4c_D^2 e^2 q^2(D-3)^2u(u-v)}{\hat{f}_{2D-3}\bar{f}_{2D-4}(2D-3)}
((\zeta_0 +\zeta_1)u v+\zeta_2v^2)\nn\\
&&>\frac{4c_D^2 e^2 q^2(D-3)^2u(u-v)}{\hat{f}_{2D-3}\bar{f}_{2D-4}(2D-3)}
(\zeta_0 +\zeta_1+\zeta_2)v^2>0,
\eea
where
\bea
&&\zeta_0=C_{4D-10}^{2D-4}C_{5D-13}^{2D-4},\nn\\
&&\zeta_1=(-2 + 3 C_{3D-7}^{2D-4}) C_{4D-10}^{2D-4} - 3 C_{5D-13}^{2D-4},\nn\\
&&\zeta_2=3 C_{3D-7}^{2D-4} - 2 C_{4D-10}^{2D-4},
\eea
and we also use the inequalities $\zeta_0>0,\zeta_0+\zeta_1>0,\zeta_0+\zeta_1+\zeta_2>0$, which are easy to check.
So, we obtain
\bea
\frac{\hat{B}^R_{2D-3}}{\hat{f}_{2D-3}}>\frac{\bar{B}^R_{2D-4}}{\bar{f}_{2D-4}}.
\eea

In the proof of Eq.\eqref{bar2d-4}, we always rewrite $C_{\ast}^{p+1}$ in $\bar{A}^R_{p+1}$ and $\bar{f}_{p+1}$ as $\frac{\ast-p}{p+1}C_{\ast}^{p}$. e.g.
\bea
\bar{f}_{p+1}&=&\frac{1}{p+1}\left(u(u+v)C_{5D-13}^{p}(5D-13-p)+(u^2+v^2)C_{4D-10}^{p}(4D-10-p)\right.\nn\\
&&\left.-3v(u+v)C_{3D-7}^{p}(3D-7-p)+2v^2C_{2D-4}^{p}(2D-4-p)\right).\nn\\
\hat{f}_{p+1}&=&\frac{1}{p+1}\left(u(u+v)C_{5D-13}^{p}(5D-13-p)+(u^2+v^2)C_{4D-10}^{p}(4D-10-p)\right.\nn\\
&&\left.-3v(u+v)C_{3D-7}^{p}(3D-7-p)\right).
\eea
From the above equations, we can see that after the rewriting, $\hat{f}_{2D-3}$ is formally equal to $\bar{f}_{2D-3}$. 
Then, according to the proof of Eq.\eqref{bar2d-4}, we have
\bea
\frac{\bar{A}^R_{2D-3}}{\hat{f}_{2D-3}}>\frac{\bar{A}^R_{2D-4}}{\bar{f}_{2D-4}}.
\eea

Thus, we have $ b'_{2D-3}> b'_{2D-4}$ and
\bea\label{sign-2d-3-2d-4}
\text{sign}(b_{p+1})\geq \text{sign}(b_{p}),~(p=2D-4).
\eea

\section{$\text{sign}(b_{p+1})\geq \text{sign}(b_{p})$, $ p=3D-9, 3D-8$}

The coefficient $b_p$ ($p=3D-7, 3D-8$) is 
\bea\label{b-3d73d8}
b_p&=&a_6 C_{6D-18}^{p}r_h^{6D-18-p}+ a_5 C_{5D-15}^{p}r_h^{5D-15-p}+a_4 C_{4D-12}^{p}r_h^{4D-12-p}\nn\\
&&+a'_5 C_{5D-13}^{p}r_h^{5D-13-p}+a'_4 C_{4D-10}^{p}r_h^{4D-10-p}+a'_3C_{3D-7}^{p}r_h^{3D-7-p}\nn\\
&=&r_h^{-p}(a_6 u^6 C_{6D-18}^{p}+ a_5 u^5 C_{5D-15}^{p}+a_4 u^4 C_{4D-12}^{p})\nn\\
&&+r_h^{-p+2}u^2(a'_5u^3 C_{5D-13}^{p}+a'_4 u^2 C_{4D-10}^{p}+a'_3uC_{3D-7}^{p}).
\eea
Define the normalized coefficients $b'_p$
\bea\label{hAp-hBp}
b'_p=\frac{r_h^p}{\hat{f}_p}b_p
=\frac{\hat{A}_p}{\hat{f}_p}+r_h^2u^2\frac{\hat{B}_p}{\hat{f}_p}
=\frac{\hat{A}^L_p}{\hat{f}_p}\l_l+\frac{\hat{A}^R_p}{\hat{f}_p}+r_h^2u^2(\frac{\hat{B}^M_p}{\hat{f}_p}(\mu^2-\o^2)+\frac{\hat{B}^R_p}{\hat{f}_p}).
\eea 
where 
\bea\label{halp-harp}
&&\hat{A}^{L}_p=2u^4\left(2u^2 C_{6D-18}^{p}+(D-7)u(u+v) C_{5D-15}^{p}-(D-5)(u^2+4uv+v^2) C_{4D-12}^{p}\right),\nn\\
&&\hat{A}^{R}_p=u^3\left((D-2)(D-4)u^3C_{6D-18}^p-3(D-2)(D-4)u^2(u+v)C_{5D-15}^{p} \right.\nn\\
&&~~~~\left.+u(( 4 D^3- 29 D^2 + 66 D-48)uv-(D^3-11 D^2+ 39 D-42)(u+v)^2)C_{4D-12}^{p}\right).\nn\\
\eea

Now let's prove  $b'_{3D-8}>b'_{3D-9}$, i.e.
\bea
&&\frac{\hat{A}^L_{3D-8}}{\hat{f}_{3D-8}}\l_l+\frac{\hat{A}^R_{3D-8}}{\hat{f}_{3D-8}}+r_h^2u^2(\frac{\hat{B}^M_{3D-8}}{\hat{f}_{3D-8}}(\mu^2-\o^2)+\frac{\hat{B}^R_{3D-8}}{\hat{f}_{3D-8}})\nn\\
&>&\frac{\bar{A}^L_{3D-9}}{\hat{f}_{3D-9}}\l_l+\frac{\bar{A}^R_{3D-9}}{\hat{f}_{3D-9}}+r_h^2u^2(\frac{\hat{B}^M_{3D-9}}{\hat{f}_{3D-9}}(\mu^2-\o^2)+\frac{\hat{B}^R_{3D-9}}{\hat{f}_{3D-9}}).
\eea
After rewriting $C_\ast^{p+1}$ as $\frac{\ast-p}{p+1}C_\ast^p$, formally, we have 
\bea
\hat{A}^{L(R)}_{3D-8}=\bar{A}^{L(R)}_{3D-8}.
\eea
Then, with the results in Sec.\eqref{sec-2d-4-3d-10}, we immediately obtain
\bea
b'_{3D-8}>b'_{3D-9},\nn
\eea
and
\bea\label{sign-3d-8-3d-9}
\text{sign}(b_{p+1})\geq \text{sign}(b_{p}),~(p=3D-9).
\eea

Then, let's prove  $b'_{3D-7}>b'_{3D-8}$, i.e.
\bea
&&\frac{\hat{A}^L_{3D-7}}{\hat{f}_{3D-7}}\l_l+\frac{\hat{A}^R_{3D-7}}{\hat{f}_{3D-7}}+r_h^2u^2(\frac{\hat{B}^M_{3D-7}}{\hat{f}_{3D-7}}(\mu^2-\o^2)+\frac{\hat{B}^R_{3D-7}}{\hat{f}_{3D-7}})\nn\\
&>&\frac{\hat{A}^L_{3D-8}}{\hat{f}_{3D-8}}\l_l+\frac{\hat{A}^R_{3D-8}}{\hat{f}_{3D-8}}+r_h^2u^2(\frac{\hat{B}^M_{3D-8}}{\hat{f}_{3D-8}}(\mu^2-\o^2)+\frac{\hat{B}^R_{3D-8}}{\hat{f}_{3D-8}}).
\eea
Following the proof of Eq.\eqref{cALp7} and taking terms including $y_2,C_{D-3}^p,C_{2D-6}^p,C_{3D-9}^p$ as zero, all the inequalities in Eqs.\eqref{bfmc55}\eqref{cALmc55}\eqref{beta012} still hold, and
we obtain
\bea
\frac{\hat{A}^L_{3D-7}}{\hat{f}_{3D-7}}>\frac{\hat{A}^L_{3D-8}}{\hat{f}_{3D-8}}.
\eea 
Taking terms including $y_2,C_{D-3},C_{2D-6},C_{3D-9}$ as zero and following the proof in Sec.\eqref{sec-car}, it is easy to deduce
\bea
\frac{\hat{A}^R_{3D-7}}{\hat{f}_{3D-7}}>\frac{\hat{A}^R_{3D-8}}{\hat{f}_{3D-8}}.
\eea 
Based on the general results in Eqs.\eqref{hBmu}\eqref{hBr}, we can get 
\bea
\frac{\hat{B}^{M(R)}_{3D-7}}{\hat{f}_{3D-7}}>\frac{\hat{B}^{M(R)}_{3D-8}}{\hat{f}_{3D-8}}.
\eea
Thus, we prove
\bea
b'_{3D-7}>b'_{3D-8},\nn
\eea
and obtain
\bea\label{sign-3d-7-3d-8}
\text{sign}(b_{p+1})\geq \text{sign}(b_{p}),~(p=3D-8).
\eea

\section{$\text{sign}(b_{p+1})\geq \text{sign}(b_{p})$, $ 3D-7\leq p\leq 4D-13$}

The coefficient $b_p$ ($3D-6\leq p\leq 4D-12$) is 
\bea\label{b-3d64d12}
b_p&=&a_6 C_{6D-18}^{p}r_h^{6D-18-p}+ a_5 C_{5D-15}^{p}r_h^{5D-15-p}+a_4 C_{4D-12}^{p}r_h^{4D-12-p}\nn\\
&&+a'_5 C_{5D-13}^{p}r_h^{5D-13-p}+a'_4 C_{4D-10}^{p}r_h^{4D-10-p}+a'_3C_{3D-7}^{p}r_h^{3D-7-p}\nn\\
&=&r_h^{-p}(a_6 u^6 C_{6D-18}^{p}+ a_5 u^5 C_{5D-15}^{p}+a_4 u^4 C_{4D-12}^{p})\nn\\
&&+r_h^{-p+2}u^2(a'_5u^3 C_{5D-13}^{p}+a'_4 u^2 C_{4D-10}^{p}).
\eea
Define the normalized coefficients $b'_p$ as
\bea\label{hAp-hBp}
b'_p=\frac{r_h^p}{f_p}b_p
=\frac{\hat{A}_p}{{f}_p}+r_h^2u^2\frac{{B}_p}{{f}_p}
=\frac{\hat{A}^L_p}{{f}_p}\l_l+\frac{\hat{A}^R_p}{{f}_p}+r_h^2u^2(\frac{{B}^M_p}{{f}_p}(\mu^2-\o^2)+\frac{{B}^R_p}{{f}_p}),
\eea 
where 
\bea\label{fp-bp}
&&f_p=u(u+v)C_{5D-13}^{p}+(u^2+v^2)C_{4D-10}^{p},\nn\\
&&{B}^M_p=2 u^2(D-3)\left(C_{5D-13}^p u (u + v) - C_{4D-10}^p (u^2 + 4 u v + v^2)\right),\nn\\
&&{B}^R_p=2 (D-3) u \left(-u{f}_p \o^2+2c_D e q u(C_{5D-13}^p u+C_{4D-10}^p(u+v))\o\right.\nn\\
&&~~~~~\left.-2c_D^2 e^2 q^2 u C_{4D-10}^p\right).\nn\\
&&\hat{A}^{L}_p=2u^4\left(2u^2 C_{6D-18}^{p}+(D-7)u(u+v) C_{5D-15}^{p}-(D-5)(u^2+4uv+v^2) C_{4D-12}^{p}\right),\nn\\
&&\hat{A}^{R}_p=u^3\left((D-2)(D-4)u^3C_{6D-18}^p-3(D-2)(D-4)u^2(u+v)C_{5D-15}^{p} \right.\nn\\
&&~~~~\left.+u(( 4 D^3- 29 D^2 + 66 D-48)uv-(D^3-11 D^2+ 39 D-42)(u+v)^2)C_{4D-12}^{p}\right).\nn\\
\eea
Following the proof of Eq.\eqref{cALp7} and taking terms including $y_2,y_3,C_{D-3}^p,C_{2D-6}^p,C_{3D-9}^p$ as zero, all the inequalities in Eqs.\eqref{bfmc55}\eqref{cALmc55}\eqref{beta012} still hold, and
we obtain
\bea
\frac{\hat{A}^L_{p+1}}{{f}_{p+1}}>\frac{\hat{A}^L_{p}}{{f}_{p}},~(3D-6\leq p\leq4D-13).
\eea 
Taking terms including $y_2,y_3,C_{D-3}^p,C_{2D-6}^p,C_{3D-9}^p$ as zero and following the proof in Sec.\eqref{sec-car}, it is easy to deduce
\bea
\frac{\hat{A}^R_{p+1}}{{f}_{p+1}}>\frac{\hat{A}^R_{p}}{{f}_{p}},~(3D-6\leq p\leq4D-13).
\eea 
Based on the proofs in Eq.\eqref{bBmu-pgeq4} and in Sec.\eqref{sec-bbr}, and taking terms including $y_2,y_3,C_{D-3}^p$,$C_{2D-6}^p,C_{3D-9}^p$ as zero, 
it is easy to deduce
\bea
\frac{{B}^{M(R)}_{p+1}}{{f}_{p+1}}>\frac{{B}^{M(R)}_{p}}{{f}_{p}},~(3D-6\leq p\leq4D-13).
\eea
Thus, we prove 
\bea
b'_{p+1}>b'_{p},~(3D-6\leq p\leq4D-13)\nn
\eea
and obtain
\bea\label{sign-3d-6-4d-13}
\text{sign}(b_{p+1})\geq \text{sign}(b_{p}),~(3D-6\leq p\leq4D-13).
\eea

After rewriting $C_\ast^{p+1}$ as $\frac{\ast-p}{p+1}C_\ast^p$, formally, we have 
\bea
f_{3D-6}=\hat{f}_{3D-6},B_{3D-6}=\hat{B}_{3D-6}.
\eea
Following the proof of $b'_{3D-7}>b'_{3D-8}$, we have
\bea
b'_{3D-6}>b'_{3D-7},\nn
\eea
and
\bea\label{sign-3d-6-3d-7}
\text{sign}(b_{p+1})\geq \text{sign}(b_{p}),~(p=3D-7).
\eea

\section{$\text{sign}(b_{p+1})\geq \text{sign}(b_{p})$, $p=4D-11,4D-12$}

The three coefficients $b_{4D-10}$,$b_{4D-11}$,$b_{4D-12}$ are
\bea
b_{4D-10}&=&r_h^{-(4D-10)}( a_6 u^6 C_{6D-18}^{4D-10}+a_5u^5 C_{5D-15}^{4D-10})\nn\\
&&+r_h^{-(4D-12)}u^2(a'_5 u^3C_{5D-13}^{4D-10}+a'_4u^2),\nn\\
b_{4D-11}&=&r_h^{-(4D-11)}( a_6 u^6 C_{6D-18}^{4D-11}+a_5u^5 C_{5D-15}^{4D-11})\nn\\
&&+r_h^{-(4D-13)}u^2(a'_5 u^3C_{5D-13}^{4D-11}+a'_4u^2C_{4D-10}^{4D-11}),\nn\\
b_{4D-12}&=&r_h^{-(4D-12)}( a_6 u^6 C_{6D-18}^{4D-12}+a_5u^5 C_{5D-15}^{4D-12}+a_4u^4)\nn\\
&&+r_h^{-(4D-14)}u^2(a'_5 u^3C_{5D-13}^{4D-12}+a'_4u^2C_{4D-10}^{4D-12}).
\eea
Define the normalized coefficients,
\bea
b'_{4D-10}&=&\frac{r_h^{4D-10}b_{4D-10}}{f_{4D-10}}=\frac{{A}_{4D-10}}{{f}_{4D-10}}+r_h^2u^2\frac{{B}_{4D-10}}{{f}_{4D-10}}\\
{b'}_{4D-11}&=&\frac{r_h^{4D-11}b_{4D-11}}{f_{4D-11}}=\frac{{A}_{4D-11}}{{f}_{4D-11}}+r_h^2u^2\frac{{B}_{4D-11}}{{f}_{4D-11}}\\
{b'}_{4D-12}&=&\frac{r_h^{4D-12}b_{4D-12}}{f_{4D-12}}=\frac{\hat{A}_{4D-12}}{{f}_{4D-12}}+r_h^2u^2\frac{{B}_{4D-12}}{{f}_{4D-12}}
\eea

After rewriting $C_\ast^{p+1}$ as $\frac{\ast-p}{p+1}C_\ast^p$, formally, we have 
\bea
A_{4D-11}=\hat{A}_{4D-11}.
\eea
Following the proof of Eq.\eqref{sign-3d-6-4d-13}, we have
\bea
b'_{4D-11}>b'_{4D-12},\nn
\eea
and
\bea\label{sign-4d-12}
\text{sign}(b_{p+1})\geq \text{sign}(b_{p}),~(p=4D-12).
\eea

Following the proof in Sec.\eqref{sec-p-d-3}, and taking terms including $y_3,y_2$, 
$C_{D-3}^p$,$C_{2D-6}^p$,\\ $C_{3D-9}^p,C_{4D-12}^p$ as zero, it is easy to check
\bea
b'_{4D-10}>b'_{4D-11},\nn
\eea
and we have
\bea\label{sign-4d-11}
\text{sign}(b_{p+1})\geq \text{sign}(b_{p}),~(p=4D-11).
\eea

\end{document}